\documentclass[]{aa}  
\usepackage{graphicx}
\usepackage{txfonts}
\usepackage{natbib}
\usepackage{subcaption}
\usepackage{epstopdf}
\begin{document} 
\title{BRITE-Constellation: Data processing and photometry\thanks{Based on data collected by the BRITE Constellation satellite mission, designed, built, launched, operated and supported by the Austrian Research Promotion Agency (FFG), the University of Vienna, the Technical University of Graz, the Canadian Space Agency (CSA), the University of Toronto Institute for Aerospace Studies (UTIAS), the Foundation for Polish Science \& Technology (FNiTP MNiSW), and National Science Centre (NCN).}}
\author{A.~Popowicz\inst{\ref{inst1}}
\and A.~Pigulski\inst{\ref{inst2}}
\and K.~Bernacki\inst{\ref{inst3}}
\and R.~Kuschnig\inst{\ref{inst4},\ref{inst5}}
\and H.~Pablo\inst{\ref{inst6},\ref{inst7}}
\and T.~Ramiaramanantsoa\inst{\ref{inst6},\ref{inst7}}
\and E.~Zoc{\l}o\'nska\inst{\ref{inst8}}
\and D.~Baade\inst{\ref{inst9}}
\and G.~Handler\inst{\ref{inst8}}
\and A.~F.~J.~Moffat\inst{\ref{inst6},\ref{inst7}}
\and G.~A.~Wade\inst{\ref{inst10}}
\and C.~Neiner\inst{\ref{inst11}}
\and S.~M.~Rucinski\inst{\ref{inst12}}
\and W.~W.~Weiss\inst{\ref{inst5}}
\and O.~Koudelka\inst{\ref{inst4}}
\and P.~Orlea\'nski\inst{\ref{inst13}}
\and A.~Schwarzenberg-Czerny\inst{\ref{inst8}}
\and K.~Zwintz\inst{\ref{inst14}}}
\institute{Instytut Automatyki, Wydzia{\l} Automatyki Elektroniki i Informatyki, Politechnika \'Sl\k{a}ska, Akademicka 16, 44-100 Gliwice, Poland\label{inst1}\\\email{apopowicz@polsl.pl}\label{inst1}
\and Instytut Astronomiczny, Uniwersytet Wroc{\l}awski, Kopernika 11, 51-622 Wroc{\l}aw, Poland \label{inst2}
\and Instytut Elektroniki, Wydzia{\l} Automatyki Elektroniki i Informatyki, Politechnika \'Sl\k{a}ska, Akademicka 16, 44-100 Gliwice, Poland\label{inst3}
\and Institute of Communication Networks and Satellite Communications, Graz University of Technology, Inffeldgasse 12, 8010 Graz, Austria\label{inst4}
\and Institut f\"ur Astrophysik, Universit\"at Wien, T\"urkenschanzstrasse 17, 1180 Wien, Austria\label{inst5}
\and D\'epartement de physique, Universit\'e de Montr\'eal, C.P.~6128, Succursale Centre-Ville, Montr\'eal, Qu\'ebec, H3C\,3J7, Canada\label{inst6}
\and Centre de recherche en astrophysique du Qu\'ebec (CRAQ), Canada\label{inst7}
\and Centrum Astronomiczne im.~M.\,Kopernika, Polska Akademia Nauk, Bartycka 18, 00-716 Warszawa, Poland\label{inst8}
\and European Organisation for Astronomical Research in the Southern Hemisphere (ESO), Karl-Schwarzschild-Str.~2, 85748 Garching, Germany\label{inst9}
\and Department of Physics, Royal Military College of Canada, PO Box 17000, Station Forces, Kingston, Ontario, K7K\,7B4, Canada\label{inst10}
\and LESIA, Observatoire de Paris, PSL Research University, CNRS, Sorbonne Universit\'es, UPMC Univ.~Paris 6, Univ.~Paris Diderot, Sorbonne Paris Cit\'e, 5 place Jules Janssen, 92195 Meudon, France\label{inst11}
\and Department of Astronomy \& Astrophysics, University of Toronto, 50 St.~George Street, Toronto, Ontario, M5S\,3H4, Canada\label{inst12}
\and Centrum Bada\'n Kosmicznych, Polska Akademia Nauk, Bartycka 18A, 00-716 Warszawa, Poland\label{inst13}
\and Institut f\"ur Astro- und Teilchenphysik, Universit\"at Innsbruck, Technikerstrasse 25/8, 6020 Innsbruck, Austria\label{inst14}
}
\date{Received; accepted}
\abstract
{The BRITE mission is a pioneering space project aimed at the long-term photometric monitoring of the brightest stars in the sky by means of a constellation of nano-satellites.
Its main advantage is high photometric accuracy and time coverage inaccessible from the ground. Its main drawback is the lack of cooling of the CCD and the absence of good shielding that would protect sensors from energetic particles.}
{The main aim of this paper is the presentation of procedures used to obtain high-precision photometry from a series of images acquired by the BRITE satellites in two modes of observing, stare and chopping.
The other aim is comparison of the photometry obtained with two different pipelines and comparison of the real scatter with expectations.}
{We developed two pipelines corresponding to the two modes of observing. They are based on aperture photometry with a constant aperture, circular for stare mode of observing and thresholded for chopping mode. 
The impulsive noise is a serious problem for observations made in the stare mode of observing and therefore in the pipeline developed for observations made in this mode, hot pixels are replaced using the information from shifted images in a series obtained during a single orbit of a satellite. In the other pipeline, the hot pixel replacement is not required because the photometry is made in difference images.}
{The assessment of the performance of both pipelines is presented. It is based on two comparisons, which use data from six runs of the UniBRITE satellite: (i) comparison of photometry obtained by both pipelines on the same data, which were partly affected by charge transfer inefficiency (CTI), (ii) comparison of real scatter with theoretical expectations. It is shown that for CTI-affected observations, the chopping pipeline provides much better photometry than the other pipeline. For other observations, the results are comparable only for data obtained shortly after switching to chopping mode. Starting from about 2.5 years in orbit, the chopping mode of observing provides significantly better photometry for UniBRITE data than the stare mode.}
{This paper shows that high-precision space photometry with low-cost nano-satellites is achievable. The proposed methods, used to obtain photometry from images affected by high impulsive noise, can be applied to data from other space missions or even to data acquired from ground-based observations.}
\maketitle
\section{Introduction}
BRITE-Constellation\footnote{BRITE stands for BRIght Target Explorer.} is a group of five nano-satellites, launched in 2013\,--\,2014, aimed at obtaining long uninterrupted time-series precision photometry in two passbands for the brightest (typically $V <$~5\,mag) stars in the sky \citep[][hereafter Paper I]{BRITE1}. This is the first space astronomy mission accomplished with nano-satellites, cubes of an edge length of 20\,cm.The scientific goals of the mission focus on variability of intrinsically luminous stars, especially those that are located at the upper main sequence, i.e., hot massive stars. They show a variety of variability types which are caused by pulsations, mass loss, rotation, stellar winds and interactions in binary and multiple systems. There are several advantages of studying bright stars. First, spectroscopic observations of such stars either exist in large number or can be easily obtained even with a small or a midsize telescope. Next, they usually have a long record of observations which in many cases allows for a study of secular variability and long-term processes. Finally, these stars are usually relatively close enough to us, so that additional information, like parallax or interferometric orbit for a binary, is available. Up to the time of writing (April 2017), photometric data for over 400 stars in seventeen fields located along the Galactic plane, were delivered. The first scientific results of the mission have been published \citep{2016A&A...588A..54W,2016A&A...588A..55P,2016A&A...588A..56B,2017MNRAS.464.2249H,2017MNRAS.467.2494P,2017arXiv170400576B,2017arXiv170401151K}.

The present paper is regarded as the third in a series devoted to the description of the technical aspects of the mission. Paper I describes the history, concept, and the main objectives of the BRITE mission, while Paper II \citep{2016PASP..128l5001P} explains design and characteristics of BRITE optics and detectors, pre-flight tests, launches, commissioning phase and in-orbit performance. In Paper II, particular attention is paid to the effects of radiation damage and ways to mitigate them. It also introduces the two modes of observing in which the BRITE satellites observe, a stare and a chopping mode. In the present paper, we give details of the two pipelines that are currently used for obtaining photometry from BRITE images and discuss the quality of the final photometry by a comparison with theoretical expectations.

\section{Data from the BRITE satellites}\label{data}
\begin{figure}[t]
\centering
\includegraphics[width=0.49\textwidth]{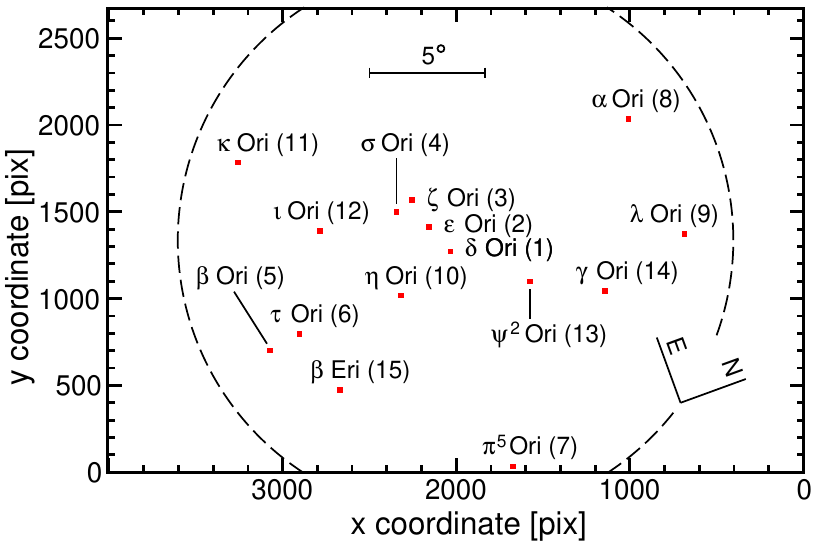}
\caption{Configuration of fifteen 32\,$\times$\,32 pixel subrasters (to scale) on the BRITE CCD detector during UBr satellite observations in the Orion I field. Stars are labeled with names and numbers indicated also in Fig.\,\ref{PSFs}. The area outside the dash-lined circle with a diameter of 24$\degr$ is affected by vignetting.}
\label{rasters_conf}
\end{figure}

\begin{figure*}[t]
\centering
\includegraphics[width=0.3\textwidth]{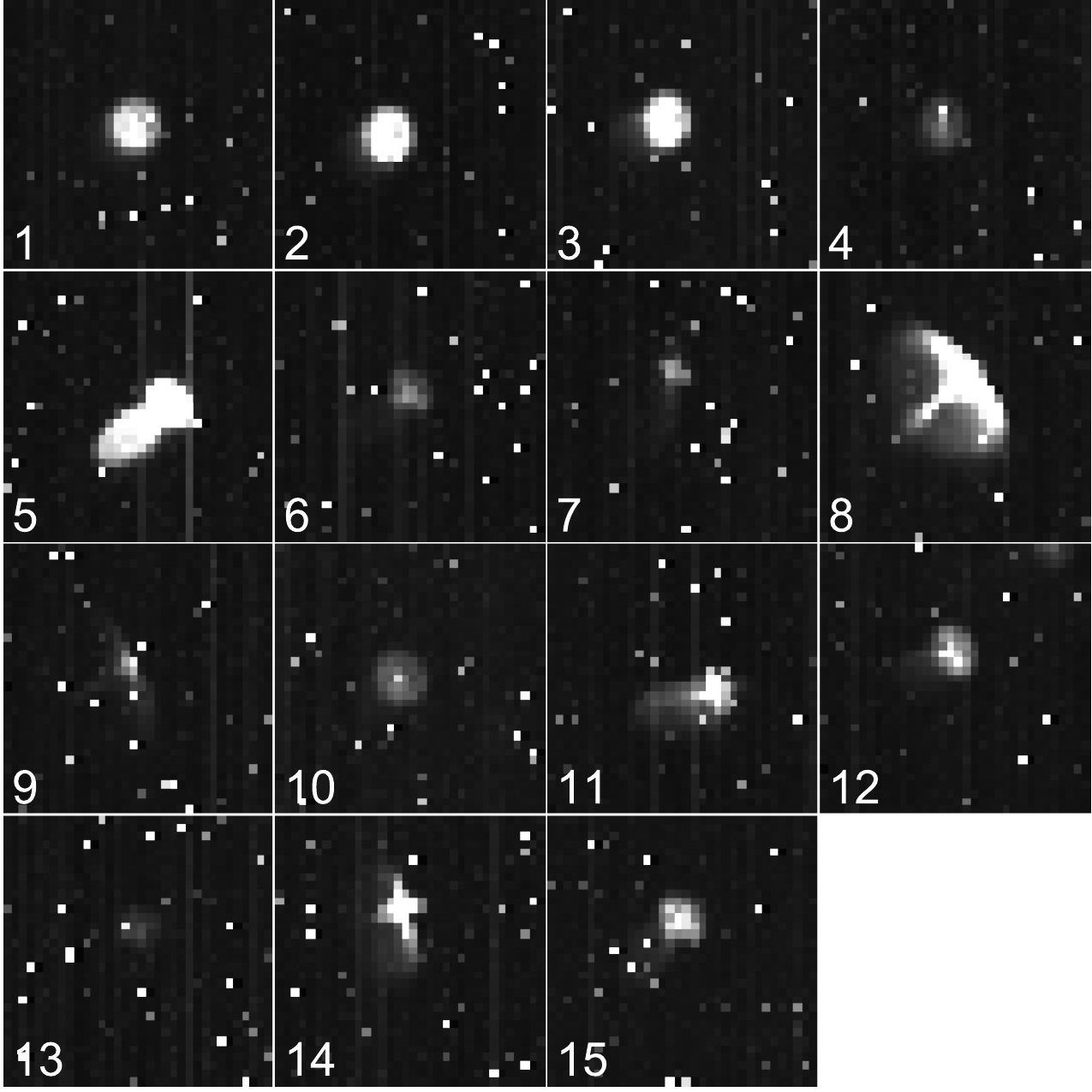}\hspace{10mm}
\includegraphics[width=0.3\textwidth]{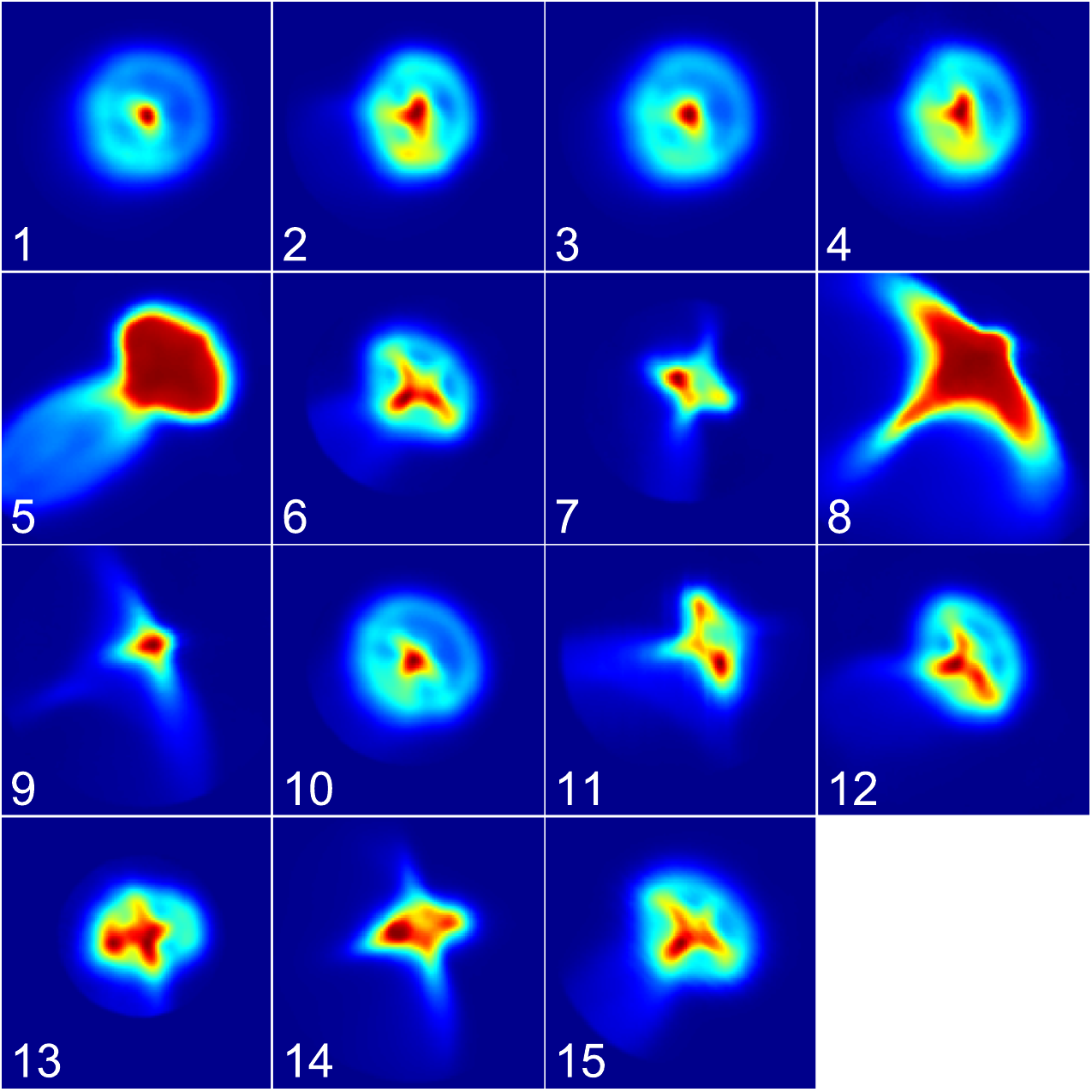}
\caption{Stellar profiles of 15 stars in Orion I field, the first field observed by the BRITE satellites. The left image shows raw subrasters, the right one, high-fidelity profile templates obtained using the Drizzle algorithm. The numbers correspond to those used to label stars in Fig.\,\ref{rasters_conf}.}
\label{PSFs}
\end{figure*}

The data downloaded from the BRITE satellites consist of pre-defined small parts of the full CCD image, which we will call subrasters throughout the paper. The subrasters are either square (in the stare mode of observing) or rectangular (in the chopping mode) having sides that range from 24 to 54 pixels depending on the star that is in it and depending on the satellite. With 27$^{\prime\prime}$ per pixel resolution, this corresponds to 11$^\prime$\,--\,24$^\prime$ in the sky. Full-frame images are downloaded only occasionally, e.g., to verify the pointing. The full field of view of the CCD covers 30$^\circ$\,$\times$\,20$^\circ$ in the sky, but is limited by vignetting to the area of about 24$^\circ$ wide and 20$^\circ$ high (Fig.\,\ref{rasters_conf}). Stars are intentionally defocussed in BRITE images. Without defocussing, a star's image would be recorded within a small fraction of a pixel, making the stellar profile undersampled and thus the photometry less precise. The price of defocussing are position-dependent and highly non-symmetric stellar profiles. The complexity of profiles increases with the distance from the image centre. A sample configuration of subrasters in the CCD plane is presented in Fig.\,\ref{rasters_conf} for the UBr satellite\footnote{As in Papers I and II, the following abbreviations for BRITE satellites will be used: BRITE-Austria (BAb), UniBRITE (UBr), BRITE-Toronto (BTr), BRITE-Lem (BLb) and BRITE-Heweliusz (BHr).} and Orion\,1 field. Stellar profiles in this field are presented in Fig.\,\ref{PSFs}. In spite of defocussing one can see that profiles sometimes show large gradients, even within a single pixel (see Fig.\,\ref{HiRes}), and therefore the images can be considered as undersampled. The high-resolution profile templates shown in the right-hand part of Fig.\,\ref{PSFs} were obtained by processing the raw frames with the filtering scheme implemented in our pipeline and then by performing resolution enhancement with the Drizzle algorithm \citep{Drizzle}. They reveal the subtle structure of the profiles. Drizzle parameters were chosen as follows: the drop size (\texttt{pixfrac}) was set to 0.2 and the output image had 10 times higher sampling than the original image.

\begin{figure*}[t]
\centering
\includegraphics[width=0.66\textwidth]{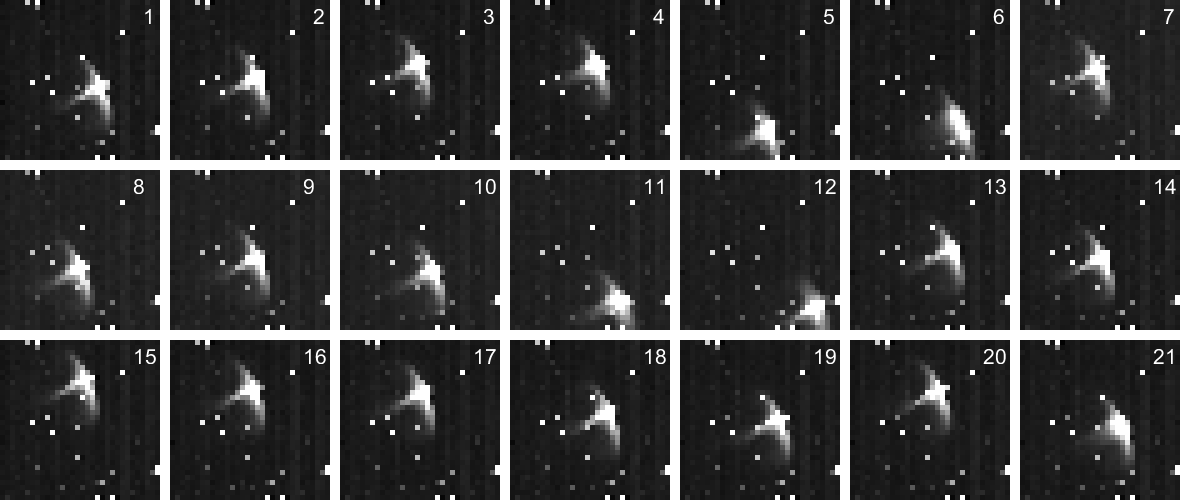}
\includegraphics[width=0.9\textwidth]{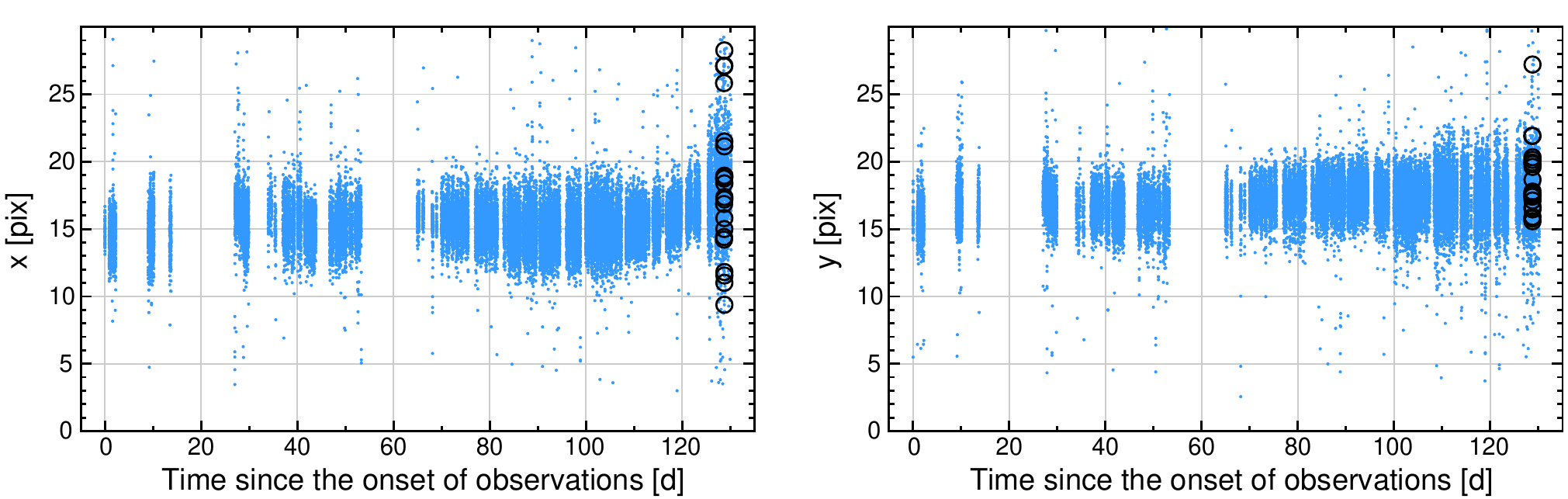}
\caption{\textit{Top:} A series of 21 consecutive 32\,$\times$\,32 pixel subrasters of $\lambda$ Ori obtained by UBr in the Orion I field. There is one blurred image in the series (6), while some are partly shifted beyond the subraster (5, 6, 11, 12). {\it Bottom:} The variations of the position of the stellar centroids in $x$ (left) and $y$ (right) coordinate for the whole run of UBr. The positions of stellar images in the presented subrasters are shown with open circles.}
\label{drifts}
\end{figure*}

The next critical problem of BRITE data is the presence of detector defects, which can be observed as distinctively brighter pixels and columns. While the former are related to the generation of dark current during exposure, the latter result from a similar process in the serial register during the readout phase. These are permanent defects induced by energetic protons hitting the CCD sensor \citep{Janesick1}. In contrast to imagers employed in ground-based observations, the dark-current generation rate in cameras located in space is unstable because of the presence of so-called random telegraph signals \citep[RTS,][]{RTS3,RTS1,RTS2}, a sudden step-like transitions between two or more metastable levels of noise that occur at random times. Thus, a standard dark frame subtraction will not work. For more details on the degradation of BRITE CCDs, the interested reader is referred to Paper II. Finally, the precision of tracking is far from perfect, resulting in occasional blurring of images or shifting the stars outside their subrasters. An example of such behaviour is presented in Fig.\,\ref{drifts}. The same figure shows also an example of tracking performance as judged by the motion of the image centroids during the first observing run of UBr.

\subsection{Modes of observing}\label{omodes}
Two years following the launch of the first BRITE satellites (UBr and BAb) in February 2013, the observations were carried out in so-called `stare' mode, in which star trackers kept satellites in a fixed orientation, so that in consecutive images stars were expected to be located at approximately the same position on the CCD, close to the centres of subrasters. An unavoidable jitter in position was also observed. For this observing mode, the pipeline presented in Sect.\,\ref{stare} was developed. 

\begin{figure}[t]
\centering
\includegraphics[width=0.48\textwidth]{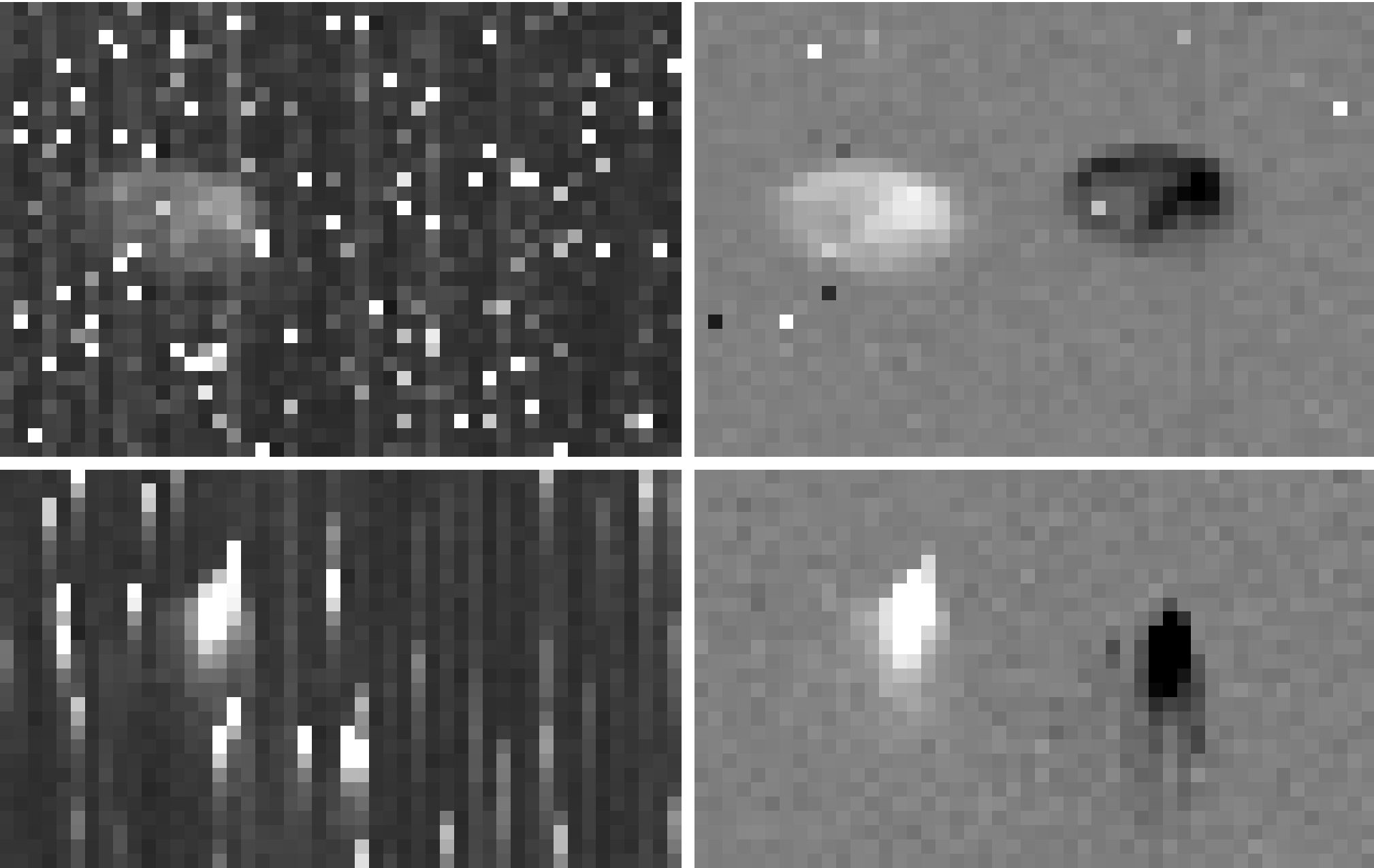}
\caption{\textit{Left:} Two examples of subrasters observed in the chopping mode, one showing extremely high noise level (\textit{top}, BAb, Cygnus I field) and the other demonstrating the CTI problem (\textit{bottom}, UBr, Scorpius I field). \textit{Right:} The corresponding difference images between two consecutive frames.}
\label{bad_images_CTI}
\end{figure}

As mentioned in Paper II, the number of hot pixels for BRITE detectors increased with time, which severely affected photometry in stare mode, especially for observations made at higher temperatures. Moreover, charge transfer inefficiency (CTI) emerged, introducing a blurring of stellar profiles and hot pixels along columns of the detector as shown in the bottom left panel of Fig.\,\ref{bad_images_CTI}. As a remedy, a new mode of observing, called `chopping' mode, was introduced. In this mode, the satellite pointing is switched between two positions separated by about 20 pixels (9$^\prime$ in the sky). In consequence, a star is placed either on the left or the right side of the subraster. After each move, a new image is taken. Subtracting consecutive frames provides difference images which are used for the photometry. Difference images contain no or very few defective pixels. Those which are left occur as a consequence of the RTS phenomenon. Sample original and difference images are presented in Fig.\,\ref{bad_images_CTI}. In chopping mode, the subrasters had to be enlarged approximately twice with respect to the stare mode, to encompass two complete stellar profiles in the difference image. Chopping mode was successfully adopted in each BRITE satellite and is currently the default mode of observing. The pipeline for chopping mode is presented in Sect.\,\ref{chop}.

\section{The pipeline for observations made in stare mode}\label{stare}
The unpredictable generation of charge in hot pixels and imperfect tracking have put high demands on the image processing pipeline. In the adopted pipeline, both the conventional image processing routines and novel approaches were implemented. Up to now, the pipeline was used to obtain light curves for over 200 stars in eight BRITE fields. 

The algorithm used in the pipeline can be divided into four main parts: classification of images (Sect.\,\ref{class}), image processing and photometry (Sect.\,\ref{photo}), compensation for intra-pixel sensitivity (Sect.\,\ref{intra}), and optimization of parameters (Sect.\,\ref{optim}). The first part aims at dividing the raw images into three types: useful, dark, and defective. In the second part, the useful images are processed in order to obtain centres of gravity (centroids) of stellar profiles and consequently to estimate stellar flux within a circular aperture. The variations of intra-pixel sensitivity are compensated in the subsequent part of the pipeline. Finally, the pipeline parameters are optimized to achieve the highest photometric precision.

\subsection{Classification of subrasters and localization of stellar images}\label{class}
The first part of the pipeline included the routines developed for finding and rejecting the images disturbed by bad tracking or data transfer problems. At this step, the dark images\footnote{There is no shutter in the optical path. Therefore, these are not real dark images but the images of  small parts of the sky devoid of bright stars. They include bias, dark, and the (negligible) stellar background.} were also identified. After the first few months of observation, it was decided that dark images would be regularly taken during each orbit by swinging the satellite, so as to move the stars outside their pre-defined subrasters. This helped to reveal the locations of hot pixels.

\begin{figure}[t]
\centering
\includegraphics[width=0.48\textwidth]{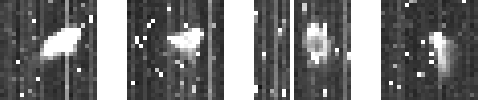}\\\vspace{0.1cm}
\includegraphics[width=0.48\textwidth]{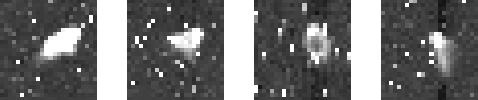}\\\vspace{0.1cm}
\includegraphics[width=0.48\textwidth]{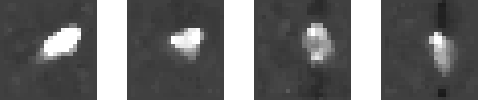}\\\vspace{0.1cm}
\includegraphics[width=0.48\textwidth]{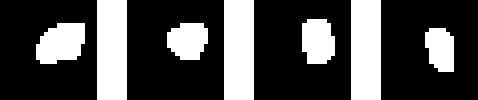}
\caption{Four consecutive steps of the first iteration of image processing. \textit{From top to bottom}: raw images; images after subtracting median values in columns; median-filtered images; stellar mask -- the final result of thresholding the stellar profile.}
\label{partA_examples}
\end{figure}

The dark current generation in the vertical charge transfer register results in columns with higher signal (see examples in the top row of Fig.\,\ref{partA_examples}). To account for this effect, the column offsets were compensated by subtracting the median intensity estimated from all pixels in a given column of the subraster. The median was used because it is highly robust with respect to the occurence of hot pixels. Since pixels containing stellar profile may also affect the median, their effect is diminished by excluding pixels which fall into stellar profile; see below. The results of this step are presented in the second row of Fig.\,\ref{partA_examples}. 

In the next step, the image is smoothed by applying a 3\,$\times$\,3 pixel median filter, in which each pixel intensity is replaced by the median of its eight neighbours. Such a procedure suppresses the hot pixels so that they do not affect the identification of the stellar region in the next step. A slight smearing of stellar profiles induced by this procedure is acceptable for the next step of defining the stellar aperture area. An example of the application of such median filtering is given in the third row of Fig.\,\ref{partA_examples}. 

The estimation of the area covered by a stellar profile (hereafter called stellar area, $A_{\rm s}$, expressed in pixels) is based on a simple thresholding of the median-filtered image obtained in the previous step. From the analysis of a few dozen stars, a value of 100~ADU was selected as a reasonable threshold. This warrants the robustness against electronic noise while keeping the ability to detect the faintest targets. In order to consistently identify the stellar profile, only the region containing the largest number of contiguous pixels is identified as a star. Smaller, isolated regions, detected during the thresholding, are rejected. This step constitutes an additional protection against identifying hot pixels or their clusters as real stars. It also discards fainter nearby stars, which are occasionally present within subrasters.

\begin{figure}[t]
\centering
\includegraphics[width=0.48\textwidth]{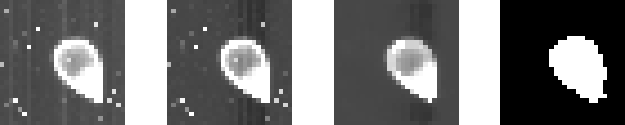}\\\vspace{0.1cm}
\includegraphics[width=0.48\textwidth]{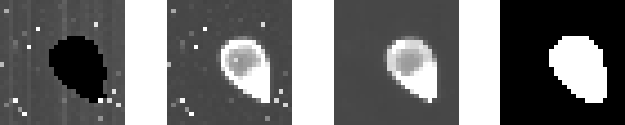}
\caption{Iterative detection of stellar area. \textit{Top row:} first iteration, \textit{bottom row:} second iteration. From left to right the same four steps as in Fig.\,\ref{partA_examples} are shown. The only difference is the leftmost image, in which the stellar area is masked as black (and ignored in calculations) in the second iteration.}
\label{iterative}
\end{figure}

Although the median is the robust measure of a column offset, it may give an overestimated value if a stellar profile extends over a large fraction of a subraster. In this case, its subtraction may result in a dimmer column (see the rightmost images in the second and third row of Fig.\,\ref{partA_examples}). As a consequence, one obtains a slightly underestimated stellar area after thresholding. In order to avoid this problem, the procedure of column subtraction and star detection is iterated. In each iteration, the median in a column is calculated ignoring the pixels included in the stellar area during the previous iteration. If two consecutive stellar areas are identical, the procedure stops and the solution is adopted. The procedure usually converges after two or three iterations. Examples of images after the first and the second iteration are shown in Fig.\,\ref{iterative}.

\begin{figure}[t]
\centering
\includegraphics[width=0.48\textwidth]{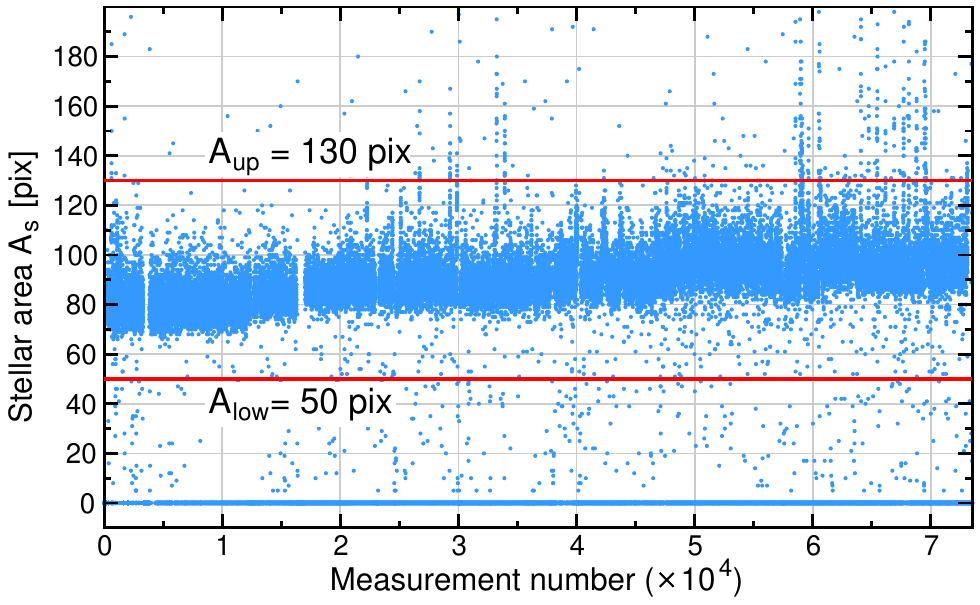}\\\vspace{0.2cm}
\includegraphics[width=0.117\textwidth]{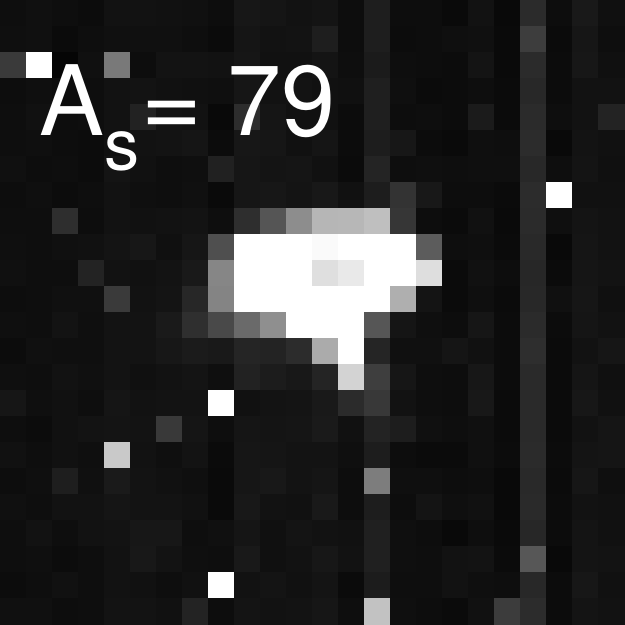}
\includegraphics[width=0.117\textwidth]{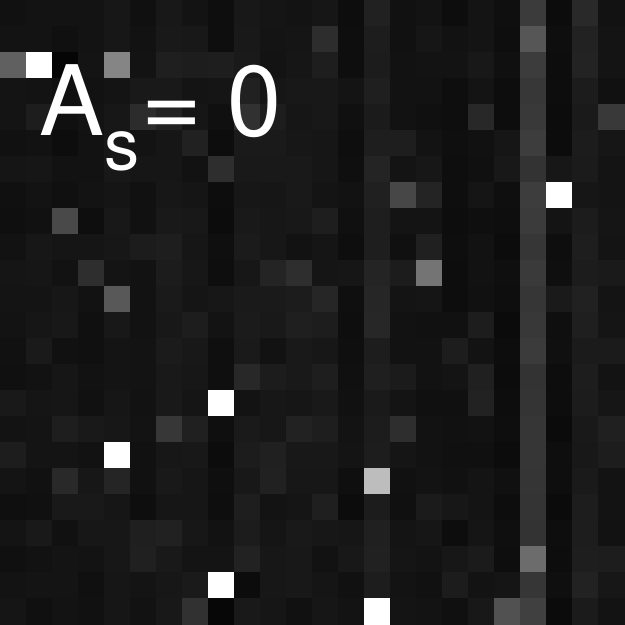}
\includegraphics[width=0.117\textwidth]{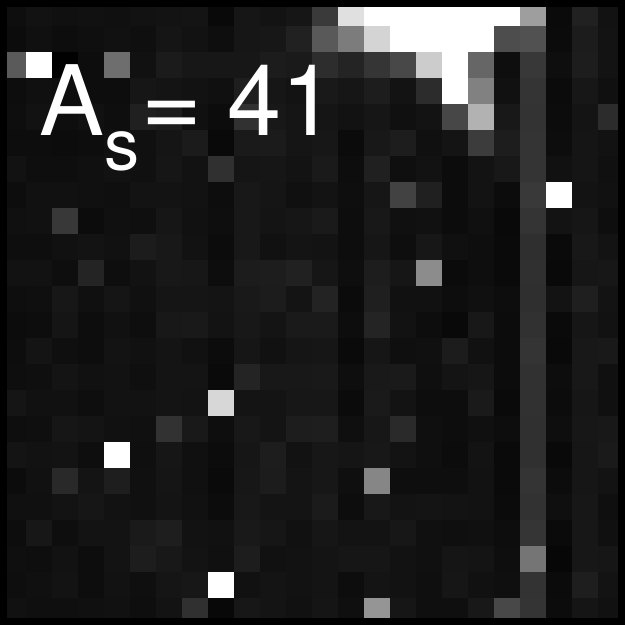}
\includegraphics[width=0.117\textwidth]{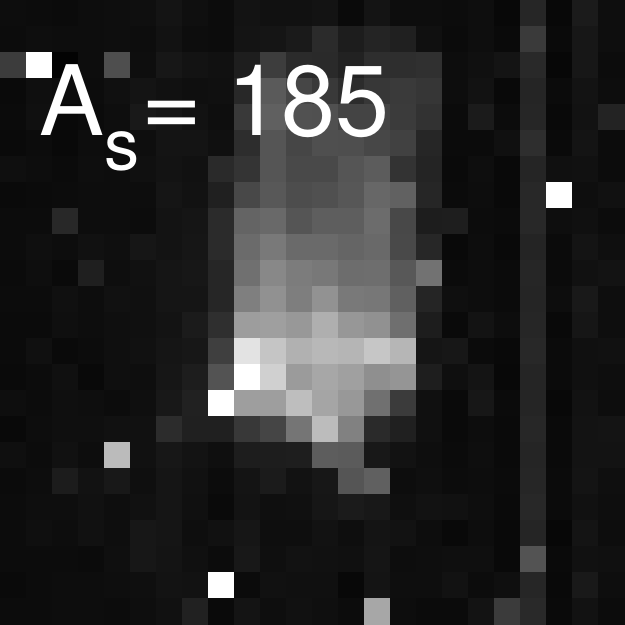}
\caption{Classification of images based on the value of $A_{\rm s}$. The upper plot shows $A_{\rm s}$ for 73\,500 consecutive measurements of HD\,127973 ($\eta$~Cen) in the Centaurus field. The two horizontal lines denote the two cut-off thresholds. The four images shown below were classified as (from left to right): useful image ($A_{\rm s} =$ 79\,pix), dark image ($A_{\rm s} =$ 0\,pix), image with stellar profile at the edge of the subraster ($A_{\rm s} =$ 41\,pix), and blurred image ($A_{\rm s} =$ 185\,pix).}
\label{classification1}
\end{figure}

Next, the values of the stellar areas for a series of images are analyzed. An example is given in Fig. \ref{classification1}. As one can see, $A_{\rm s}$ clusters around 80\,--\,100 pixels with some outliers on both sides. The $A_{\rm s}$ markedly larger than the mean come from blurred images. Those with $A_{\rm s}$ significantly smaller than the mean appear due to the stellar profile located at the subraster edge. Finally, the images for which $A_{\rm s} =$ 0, i.e.~no stellar area was detected, are good candidates for dark frames. The final step in classification is setting two threshold values, lower, $A_{\rm low}$, and upper, $A_{\rm up}$. The images with $A_{\rm s} =$ 0 are classified as dark, those with $A_{\rm low} \leq A_{\rm s} \leq A_{\rm up}$ as useful, and those with 0 $< A_{\rm s} < A_{\rm low}$ and $A_{\rm s} > A_{\rm up}$ are marked as defective and rejected. Several methods, e.g.~based on $\sigma$-clipping or local standard deviation, were tested to automatically set the two thresholds. Unfortunately, the values of $A_{\rm s}$ may occasionally fluctuate due to the intrinsic variability and temperature-dependent changes of the size of the stellar profile. Therefore, it was decided that for each star the thresholds will be set manually. They were chosen carefully with relatively large margins to prevent excluding possible short-term intrinsic brightenings or dimmings of a star.
\begin{figure*}[t]
\centering
\includegraphics[width=0.85\textwidth]{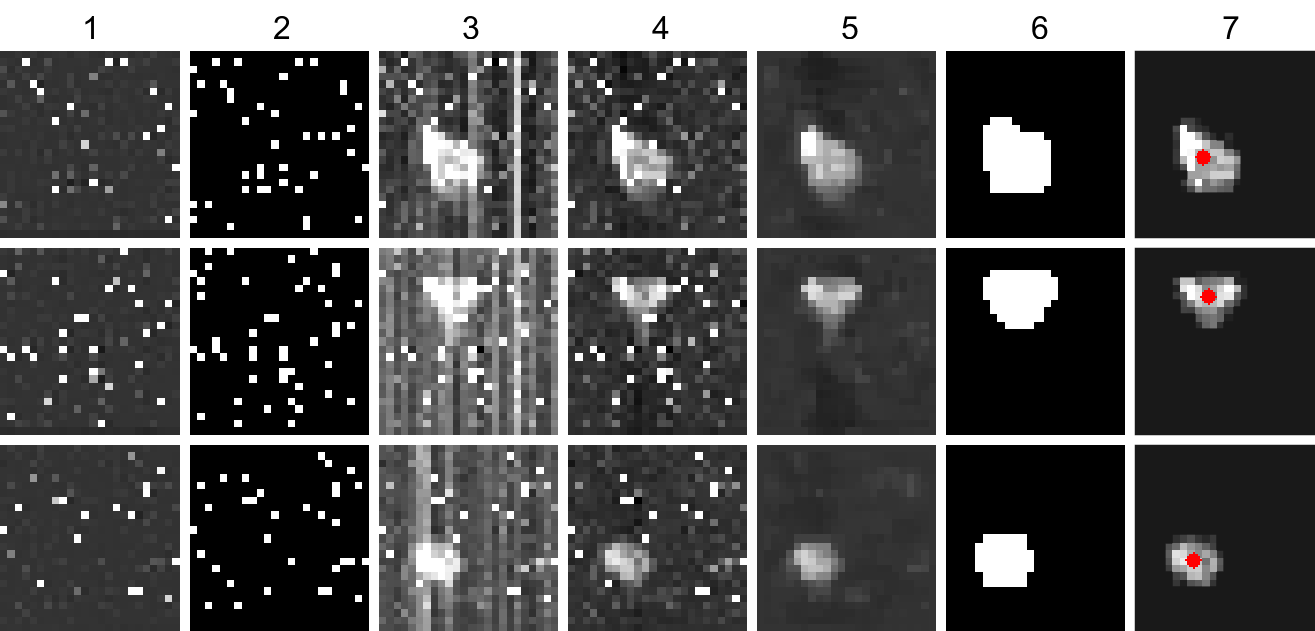}
\caption{Seven steps of initial estimation of the position of a centroid for three sample subrasters. From left to right: (1) master dark frame, (2) mask of bad pixels (thresholded master dark frame), (3) raw subraster, (4) subraster after subtracting median column offset (first iteration), (5) the same subraster after median filtering (first iteration), (6) final stellar mask, (7) $I_{\rm col}$ subraster (see text) with the position of the centroid marked by a red dot.}
\label{centr_image}
\end{figure*}

\subsection{Image processing and photometry}\label{photo}
The next part of the pipeline is aimed at obtaining aperture photometry of a star from a series of images. Only images classified previously as useful are considered. The dark frames are median-averaged to create a master dark frame, in which the pixels are classified as hot if their intensity is above a given threshold, $S_{\rm bad}$. As a result of this thresholding, a mask of bad pixels is obtained; see the second column in Fig.\,\ref{centr_image}. The $S_{\rm bad}$ parameter is optimized in the last part of the pipeline (Sect.\,\ref{optim}). 

At this stage, the part of a subraster corresponding to the stellar area (let us call it the stellar mask) is derived in the same iterative way as was performed during the classification (columns 3 to 6 in Fig.\,\ref{centr_image}). Once this is done, the image corrected for column offsets is multiplied by the stellar mask and each hot pixel within the mask is replaced by the median intensity of the neighbouring pixels. The resulting image will be referred to as $I_{\rm col}$ (last column in Fig. \ref{centr_image}). Then, the initial coordinates of the stellar centroid ($x_{\rm centr}$, $y_{\rm centr}$) are calculated as a centre of gravity from the following equations:
\begin{equation}
x_{\rm centr}= \dfrac{\sum\limits_{x=1}^{X} \sum\limits_{y=1}^{Y}x I_{\rm col}(x,y)}{\sum\limits_{x=1}^{X} \sum\limits_{y=1}^{Y}I_{\rm col}(x,y)},\quad y_{\rm centr}= \dfrac{\sum\limits_{x=1}^{X} \sum\limits_{y=1}^{Y}y I_{\rm col}(x,y)}{\sum\limits_{x=1}^{X} \sum\limits_{y=1}^{Y}I_{\rm col}(x,y)},
\label{COG1}
\end{equation}
where $X$ and $Y$ are the widths of the $I_{\rm col}$ image, in the $x$ and $y$ coordinate, respectively. 

A precise estimation and replacement of the signal in hot pixels is a crucial part of the pipeline. For the purpose of image classification and initial estimation of the stellar centre, the hot pixels were replaced by a simple median of intensities of eight neighbouring pixels. Although this is sufficient for the classification and initial determination of the stellar centre, the final photometry requires a more sophisticated filtering scheme. Multiple interpolation approaches utilizing the intensity estimation based on a pixel neighbourhood have been tested \citep[e.g.][]{CWM,nonlinearfilters,LAcosmic,Popowicz1}. Unfortunately, due to the significant undersampling of BRITE images, the fitting algorithms produced unsatisfactory results. This is why methods based on the neighbouring pixels were abandoned in favour of the approach based on the analysis of a series of exposures obtained during the same orbit.

The method we propose utilizes the fact that during a single orbit typically more than 40 frames are taken. Imperfect tracking causes considerable changes of the position of a star within a subraster (see Fig.\,\ref{drifts}). Surprisingly, this can help in obtaining a good estimation of the signal, which can be used to replace the one in a hot pixel. The adopted procedure of hot pixel interpolation is the following. Let us consider that we want to perform corrections for hot pixels in the $k$-th subraster, $I_k$, of a series of $N$ subrasters taken during a single orbit. For this purpose, the $I_{\rm col}$ subrasters are used. Let the centroid calculated with Eq.\,(\ref{COG1}) be equal to $(x_k,y_k)$ for that subraster. Let us now denote the centroids of the other images in the series as $(x_i,y_i)$,  $i = \mbox{1}, ..., N$. The differential centroids $(\Delta x_i , \Delta y_i )$ can now be defined as follows:
\begin{equation}
\left\{\begin{aligned}
\Delta x_i &= x_i-x_k,\\
\Delta y_i &= y_i-y_k,\quad i = \mbox{1}, ..., N.
\end{aligned}
\right.
\end{equation}
Let us also define fractional shifts $(\delta x_i, \delta y_i)$:
\begin{equation}
\left\{\begin{aligned}
 \delta x_i &= \Delta x_i - \texttt{round}(\Delta x_i),\\ 
 \delta y_i &= \Delta y_i - \texttt{round}(\Delta y_i),\quad i = \mbox{1}, ..., N,
\end{aligned}\right.
\end{equation}
where \texttt{round} stands for rounding a number to the nearest integer. By definition, fractional shifts range between $-$0.5 and 0.5. When in two subrasters $\delta x_i$ and $\delta y_i$ are close to each other, we may assume that in these subrasters stellar profiles are sampled in a similar way. Since stellar profiles are in general under-sampled in BRITE images, the fractional shifts play a very important role, especially when the intensities are to be replaced using other subrasters. The idea of the method is to replace a hot pixel using only subrasters shifted with respect to the considered one by an amount close to an integer number of pixels in both directions. This means that the replacement procedure will involve only those stellar profiles which are sampled in a similar way. Adopting this procedure leads us to the following formula for the signal that will be used to replace a hot pixel at the position $(x_h,y_h)$ in $I_k$ ($\widetilde{I_k}$ is the $I_k$ subraster with hot pixels corrected):
\begin{equation}\label{HPweighting}
\begin{aligned}
\widetilde{I_k}(x_h,y_h) &= \dfrac{\sum\limits_{i=1, i\neq k }^N w_i I_i(x_h+\Delta x_i, y_h+\Delta y_i)}{\sum\limits_{i=1}^N w_i},\\
w_i &= \dfrac{1}{\sqrt{\delta x_i^2+\delta y_i^2}}.
\end{aligned}
\end{equation}
The pixels flagged as hot in the thresholded master dark frame are excluded from the sum in Eq.\,(\ref{HPweighting}).

By trial and error we came to the conclusion that it is optimal to use only about ten subrasters with the highest weights in the sum in Eq.\,(\ref{HPweighting}). Using a larger number of subrasters leads to a decrease of the photometric precision because more subrasters with larger fractional shifts (albeit with smaller weights) contribute to the sum. On the other hand, reducing the number of summed pixels would lead to an increase of noise. As a compromise, $N=$ 10 was used.

Once the intensity of a hot pixel is replaced with the value calculated from Eq.\,(\ref{HPweighting}), the centre of gravity of the stellar profile may change. Therefore, the procedure of replacement of hot pixels was iterated. In general, there was no recognizable improvement of the light curve quality after the second iteration, so that it was assumed that two iterations are sufficient.

The presented approach is based on the assumption that stellar flux does not change significantly during a single-orbit observation (10\,--\,50 minutes). This assumption is well justified for most of the observed stars. However, even if this is not the case, the presented procedure introduces only a very small bias, which has a negligible effect on the final light curve. This is because the number of hot pixels in a given subraster is small in comparison with the number of pixels falling into the aperture containing the stellar profile.

In the last step, circular aperture photometry with the radius optimized as explained in Sect.\,\ref{optim} is performed, wherein only final, filtered subrasters $\widetilde{I_k}$ are used. The intensities of all the pixels, whose positions are within the aperture, are summed. 

\subsection{Compensation for intra-pixel sensitivity variations}\label{intra}
\begin{figure}[t]
\centering
\includegraphics[width=0.48\textwidth]{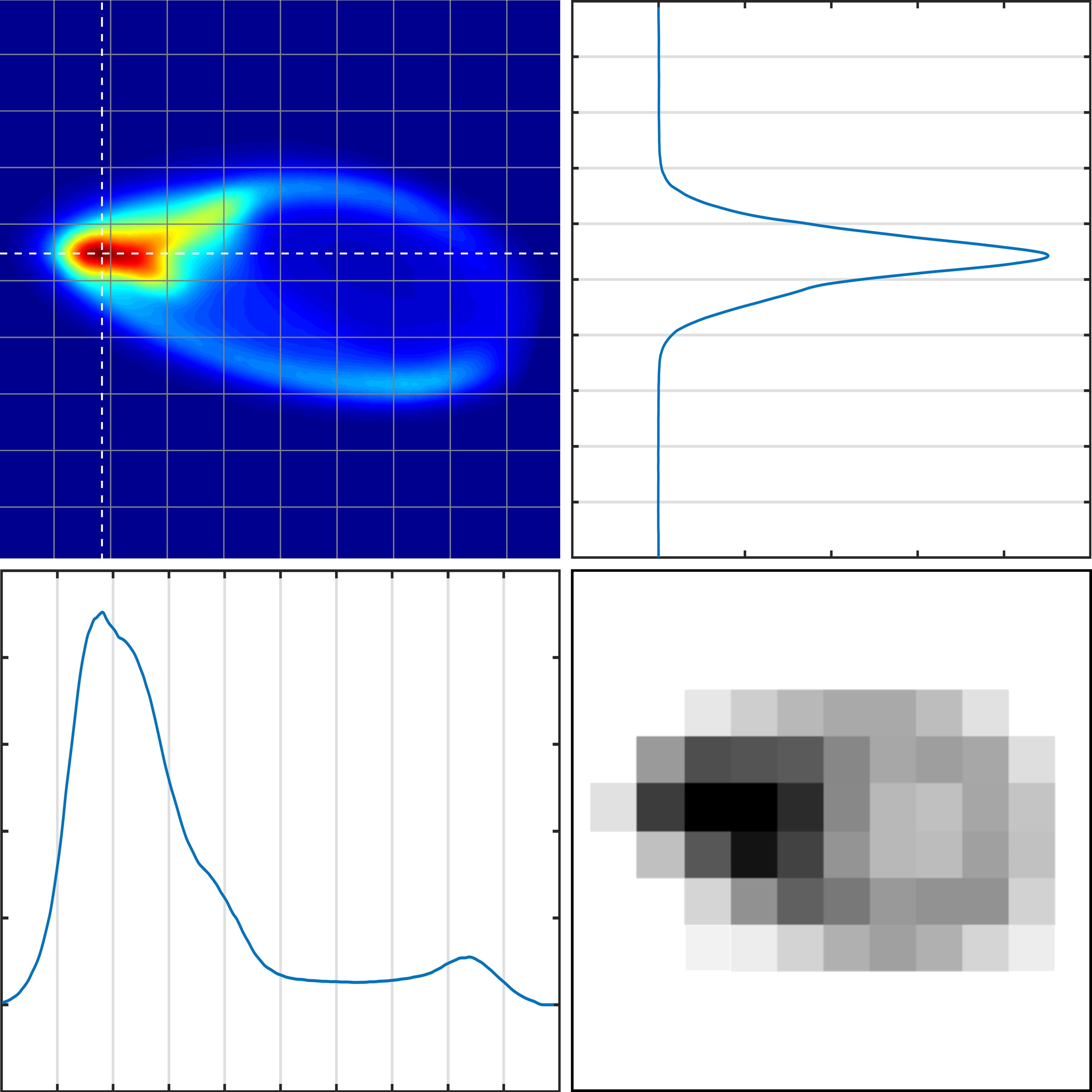}
\caption{An example of a complex stellar profile in BRITE observations. \textit{Top left:} high-resolution template obtained with Drizzle algorithm, \textit{top right:} cross section in $y$ coordinate, \textit{bottom-left:} cross section in $x$ coordinate, \textit{bottom-right:} single frame example. The positions of cross sections are indicated in the top left image with the dashed lines. Grid lines are separated by one pixel.}
\label{HiRes}
\end{figure}

It is well known that sensitivity of a pixel varies slightly across its surface \citep{intrapix1,intrapix2}. Usually, the sensitivity is the highest in the pixel centre, decreasing towards its edges. Since stellar profiles in BRITE images have complex shapes, it was expected that the variations of intra-pixel sensitivity would noticeably affect the photometry. In Fig.\,\ref{HiRes}, a sample high-resolution stellar profile is shown. It was obtained by averaging  1000 interpolated low-resolution images, an example of which is depicted in the right bottom image of Fig.\,\ref{HiRes}. As one can clearly see, there are significant intensity gradients within some pixels, and it is highly probable that the photometry will be affected by the sub-pixel shifts of a stellar profile.

\begin{figure}[!t]
\centering
\includegraphics[width=0.45\textwidth]{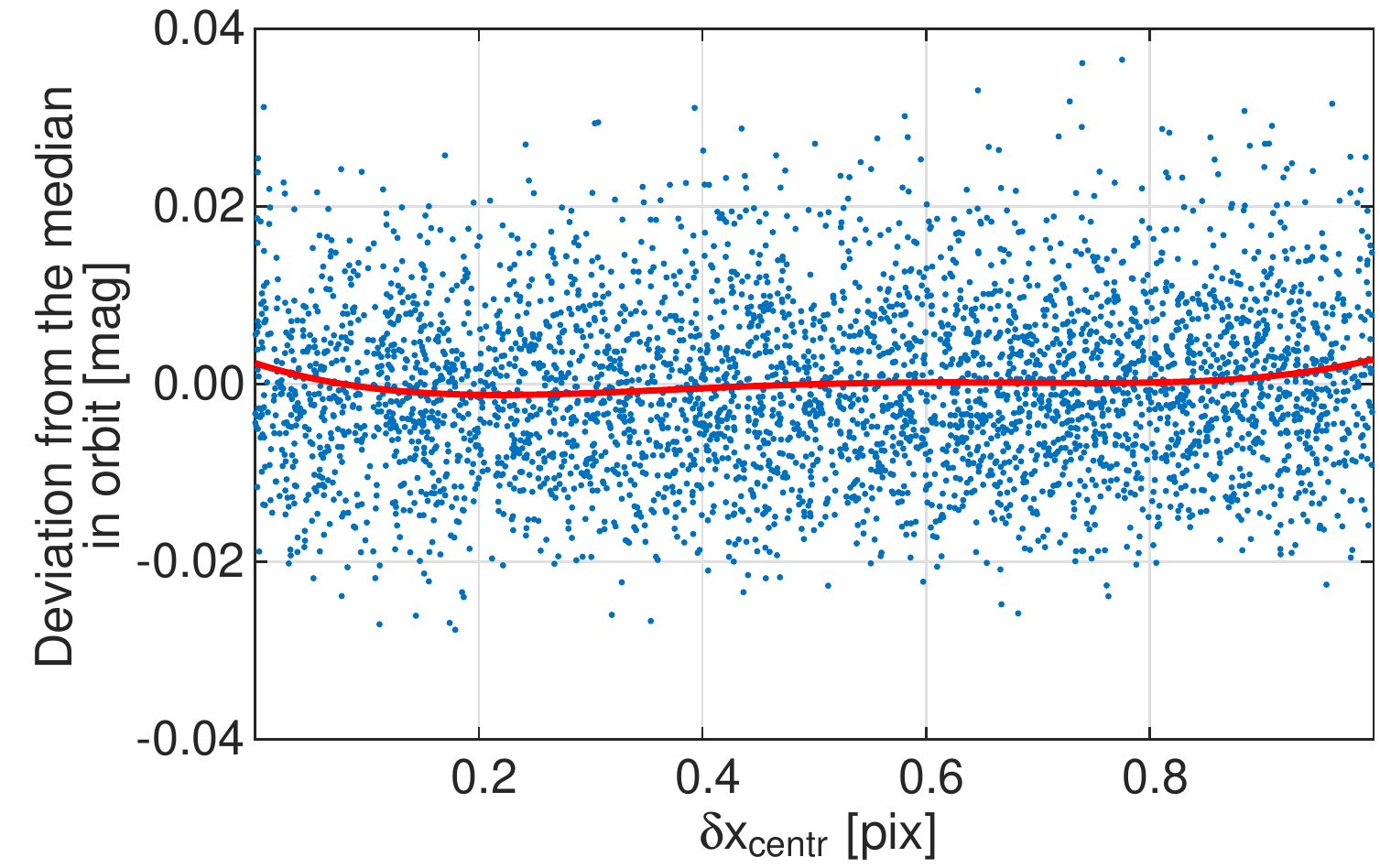}\\
\includegraphics[width=0.45\textwidth]{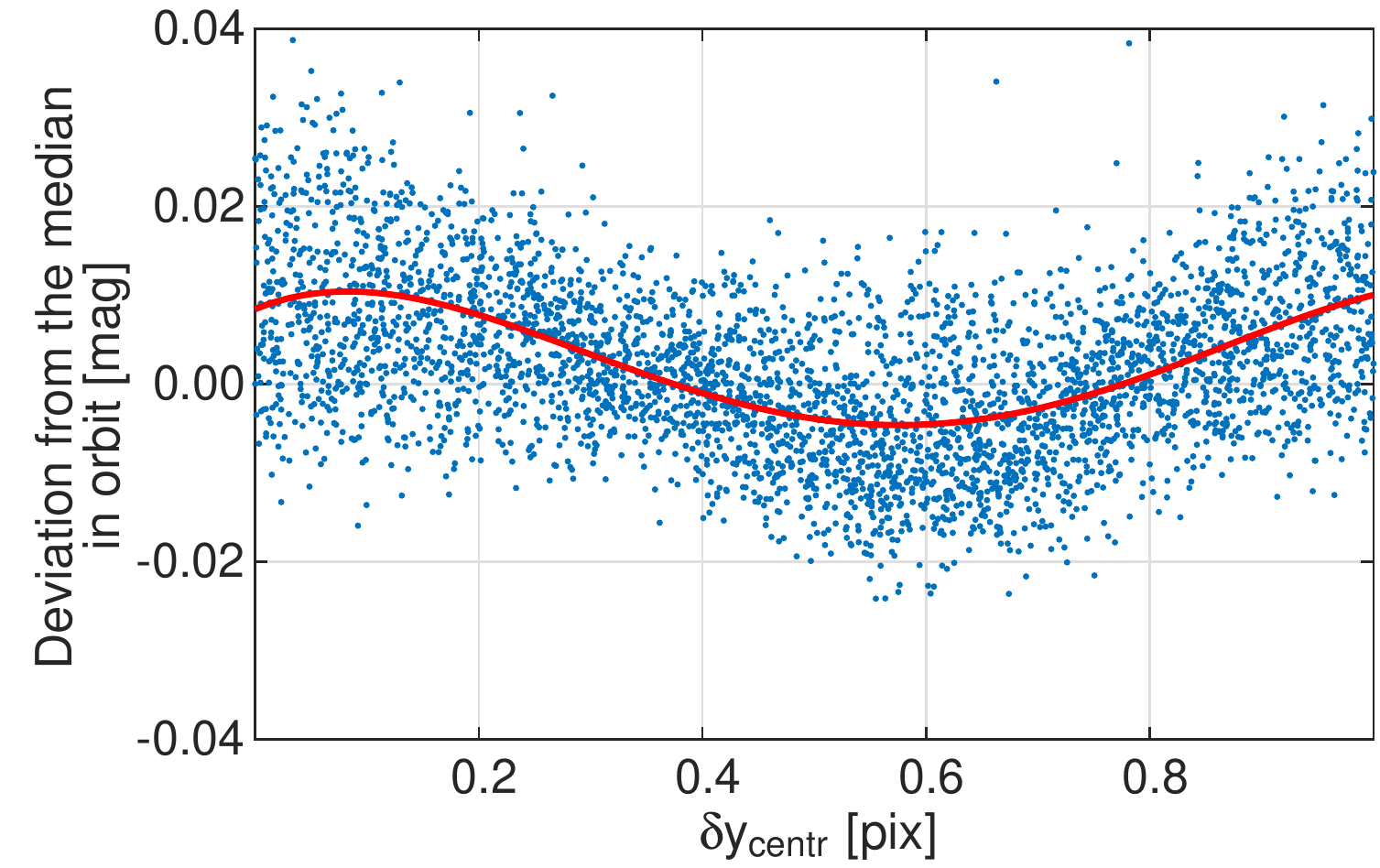}
\caption{Intra-pixel variations and the fitted fourth-order polynomials (red lines) in $x$ (top) and $y$ (bottom) coordinates. These plots were created for the same star as in Fig.\,\ref{HiRes}.}
\label{intra_pix}
\end{figure}

In order to assess and then compensate for the intra-pixel sensitivity variations, the following procedure was performed for each star. First, the median stellar flux in each orbit was computed. Then, for each measurement the deviation from the median was plotted against the fractional centroid position $(\delta x_{\rm centr}, \delta y_{\rm centr})$, defined as follows:
\begin{equation}
\left\{\begin{aligned}
 \delta x_{\rm centr} &= \texttt{frac}(x_{\rm centr}) \equiv x_{\rm centr} - {\lfloor}x_{\rm centr}{\rfloor},\\ 
 \delta y_{\rm centr} &=  \texttt{frac}(y_{\rm centr}) \equiv y_{\rm centr} - {\lfloor}y_{\rm centr}{\rfloor},
\end{aligned}\right.
\end{equation} 
where $\lfloor\,\rfloor$ is the floor function. The dependencies are approximated by a fourth-order polynomial. An example of such a fit is shown in Fig.\,\ref{intra_pix}. The larger effect in the $y$ coordinate is not a surprise because there are larger gradients in the cross section along this coordinate in comparison with the $x$ coordinate (Fig.\,\ref{HiRes}). The fitted polynomials are used in the next step to compensate for the intra-pixel variability. This is the only correction to the derived fluxes implemented in this pipeline. Intra-pixel corrections account only for the effect averaged over all pixels falling in the aperture. In other words, it is assumed that dependency between the fractional centroid position and the photometric offset is the same for the whole subraster. In reality, this is not always the case because stellar profiles change due to the temperature effect and smearing. In consequence, intra-pixel correction accounts only for a part of the position-dependent instrumental effect. It was decided that the remaining instrumental effects (such as possible flux dependencies on CCD temperature, centroid position, etc.) will be accounted independently of the photometric pipeline by the end-user. Examples of the required corrections are shown e.g.~by \cite{2016A&A...588A..55P} and \cite{2017arXiv170400576B}. Decorrelations with position would account for most of the residual dependencies provided that the correlation function is relatively smooth. A combination of both corrections, one implemented in the pipeline, the other accounted by decorrelations, seems to be the best solution when dealing with position-dependent instrumental effects.

\subsection{Optimization}\label{optim}
The last part of the algorithm is actually the loop which involves procedures described in Sects \ref{photo} and \ref{intra}, run for a set of different hot pixel thresholds $S_{\rm bad}$ and various aperture radii $R$. In our estimation of quality, we used the median absolute deviation (MAD) calculated for all $N_{\rm orb}$ orbits. The quality parameter, $Q$, is the median of normalized MADs, defined as follows:
\begin{equation}
Q = \textrm{median} \left\{ {\frac{\textrm{MAD}_i}{\sqrt{N_i}}} \right \},\quad i = 1, ..., N_{\rm orb},
\label{Qm}
\end{equation}
\begin{equation}
\textnormal{MAD}_i = \textnormal{median}\left\{|J_{ki} - \textnormal{median}({\mathbf{J_i}})|\right\}, \quad k = 1, ..., N_i,
\end{equation}
where $J_{ki}$ is $k$-th measured stellar flux in the $i$-th orbit, $N_i$ is the number of points in the $i$-th orbit and $\mathbf{J_i}$ is the set of all measurements in $i$-th orbit. Using median-based indicators allowed us to reduce the influence of outliers within orbits and the worst orbits, corrupted e.g.~by stray light, on the value of $Q$ (the median of MADs for all orbits is used in its definition, see Eq.~(\ref{Qm})). The normalization by $\sqrt{N_i}$ reflects the noise reduction due to the averaging of $N_i$ measurements in the $i$-th orbit. The values of $N_i$ range between several and over 160 depending on the satellite and field. The number of orbits $N_{\rm orb}$ per setup (see Sect.\,\ref{normvschop} for the definition of setups) has a wide range and amounts from about a dozen up to almost 1500.

\begin{figure}[t]
\centering
\includegraphics[width=\linewidth]{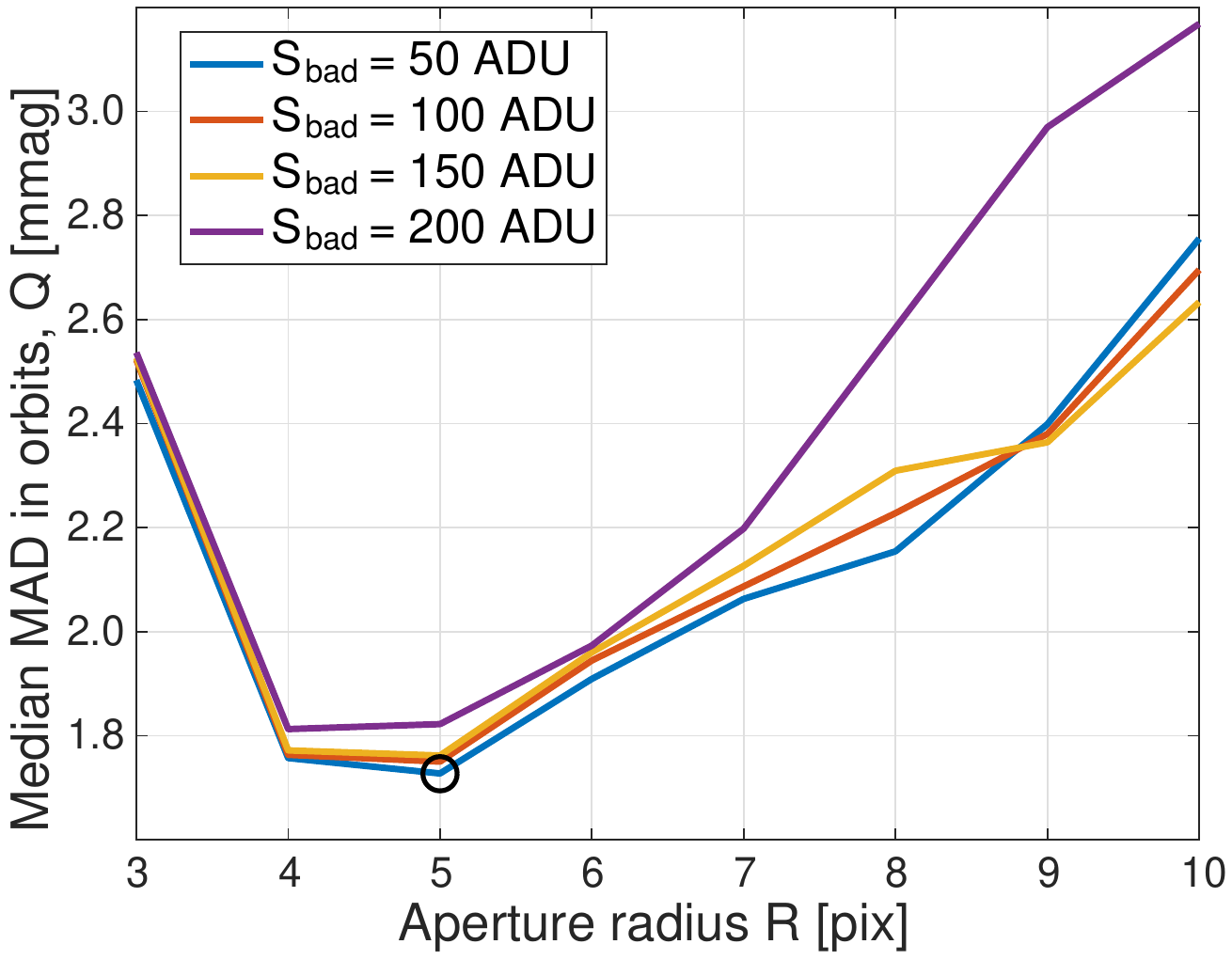}
\caption{The optimization curves, $Q$ vs.~aperture radius, $R$ (in pixels), for a sample data set (UBr observations of HD\,31237 = $\pi^5$\,Ori in the Ori I field). The optimal parameters,  $S_{\rm bad}=$ 50\,ADU, $R=$ 5 pixels, are marked by a black circle.}
\label{optimcurve}
\end{figure}

The optimization was performed for each star and setup combination independently. It was carried out for a set of four values of $S_{\rm bad}$, $S_{\rm bad}\in$ \{50, 100, 150, 200\}~ADU and eight values of $R$, $R\in$ \{3, 4, 5, 6, 7, 8, 9, 10\} pixels. The adopted range always encompassed the minimum $Q$ value. An example of optimization is presented in Fig.\,\ref{optimcurve}. 

The pipeline described in this section was applied to all data obtained in the stare mode of observing, i.e.~the first eight fields observed by the BRITE satellites (see Table \ref{rel-flds}) and to the test data in the Perseus field obtained in the chopping mode. The photometry made with this pipeline is now available as Data Release 2; see Appendix \ref{drs} for the explanation of data releases.

\section{The pipeline for observations made in the chopping mode}\label{chop}
As explained in Sect.\,\ref{omodes}, the chopping mode was introduced as a remedy for the appearance of CTI and the increasing number of hot pixels in BRITE detectors. The photometry for images taken in this mode is carried out with difference images (Fig.\,\ref{bad_images_CTI}). A difference image is defined as an image obtained by a subtraction of two consecutive images: $I_k-I_{k+1}$. In order to secure the same number of difference images as in the stare mode, each image is used twice, i.e., the next difference image is $I_{k+1}-I_{k+2}$. For the last image in each orbit the order of subtraction is reversed: $I_{\rm last}-I_{{\rm last}-1}$. The pipeline developed for the chopping mode is much less complex than that proposed for stare mode. A set of adaptive apertures was introduced to better fit the stellar profiles. The algorithm for chopping mode can be divided into two main parts: fully automatic classification and photometry with both intra-pixel sensitivity compensation and optimization.

\subsection{Automatic classification}\label{autclass}
\begin{figure}[!t]
\centering
\includegraphics[width=0.23\textwidth]{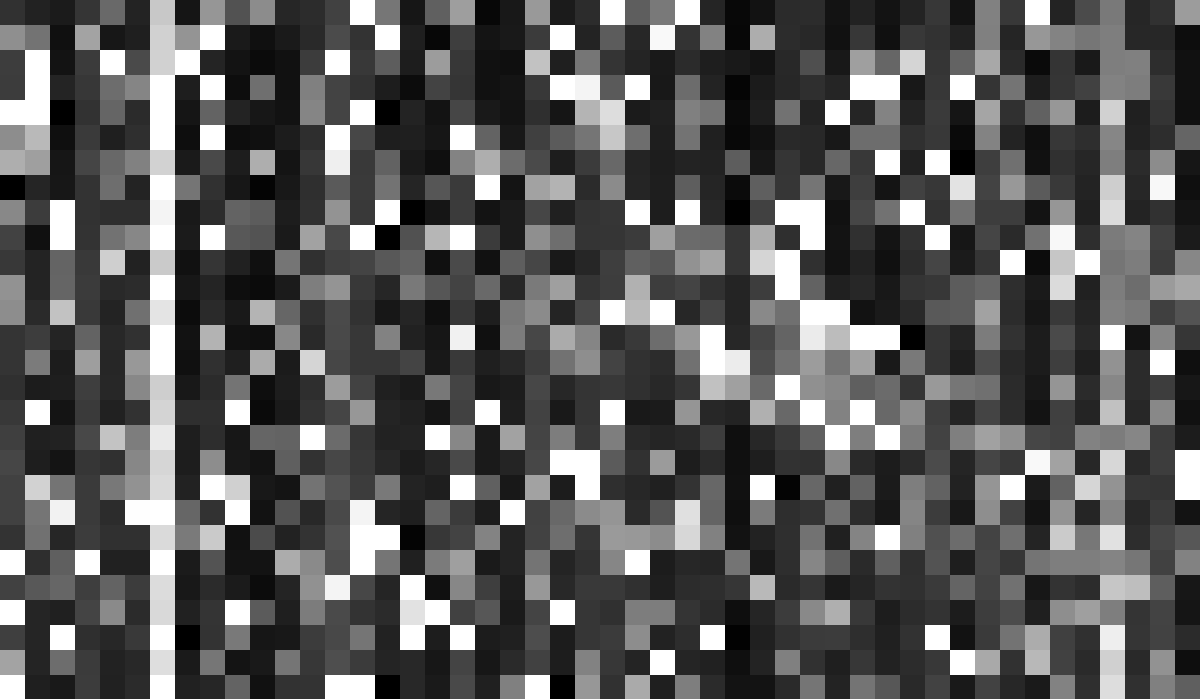}
\includegraphics[width=0.23\textwidth]{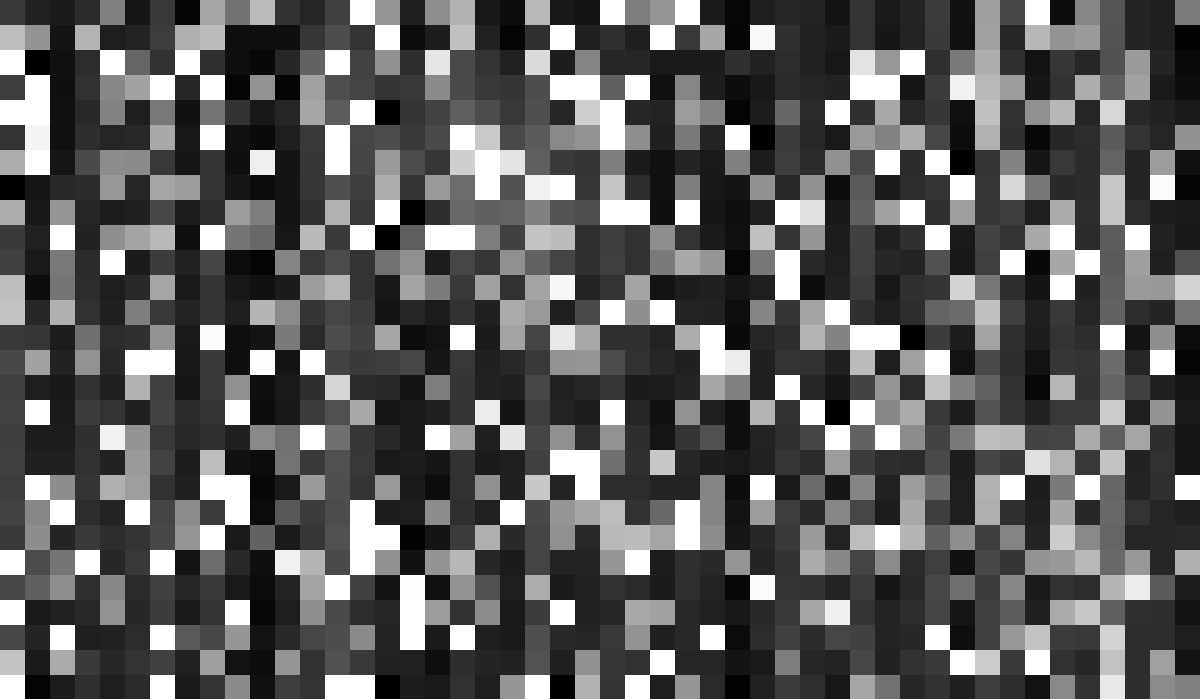}\\\vspace{0.07cm}
\includegraphics[width=0.23\textwidth]{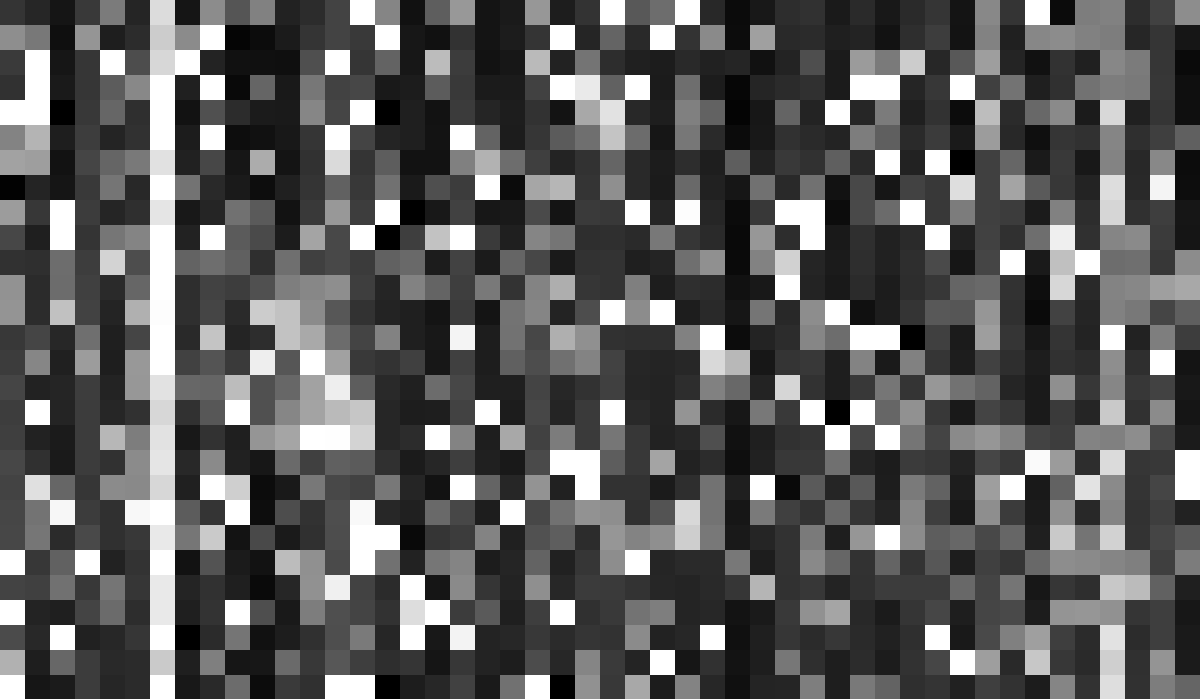}
\includegraphics[width=0.23\textwidth]{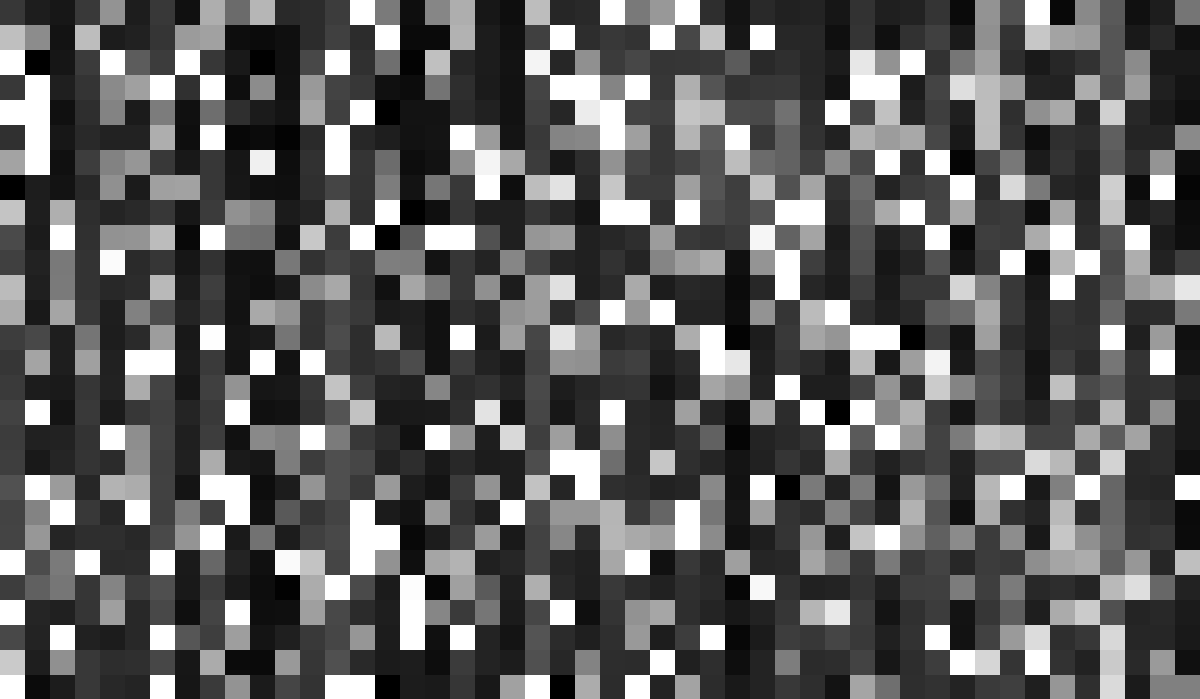}\\\vspace{0.07cm}
\includegraphics[width=0.23\textwidth]{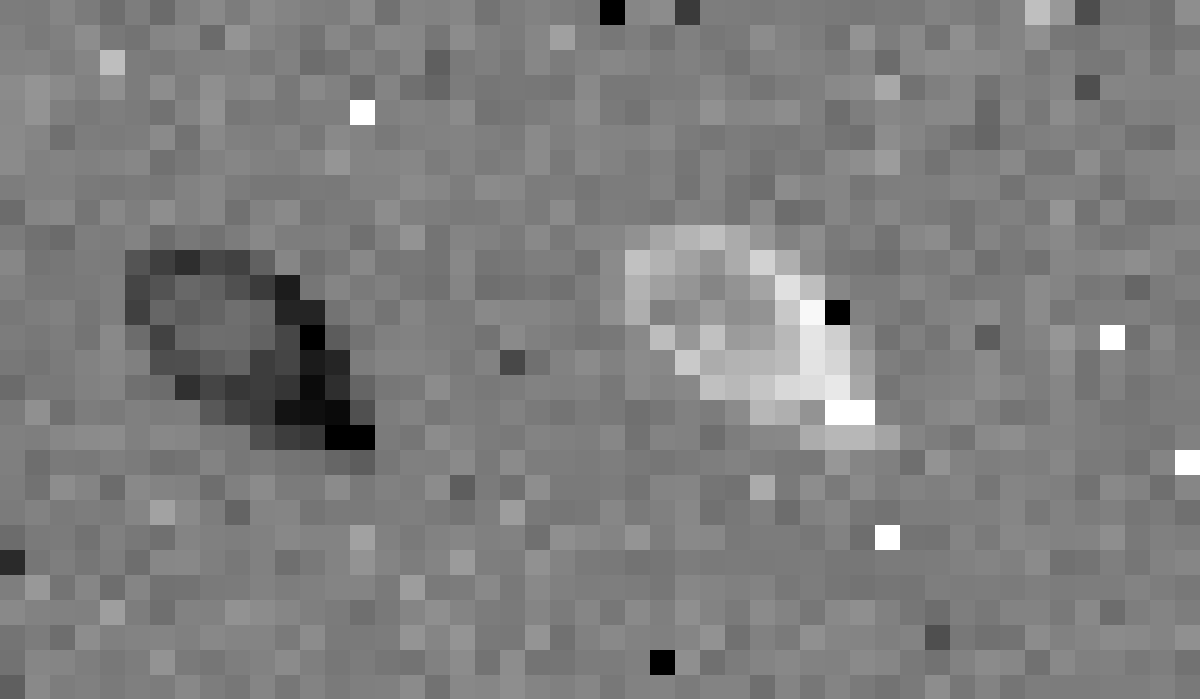}
\includegraphics[width=0.23\textwidth]{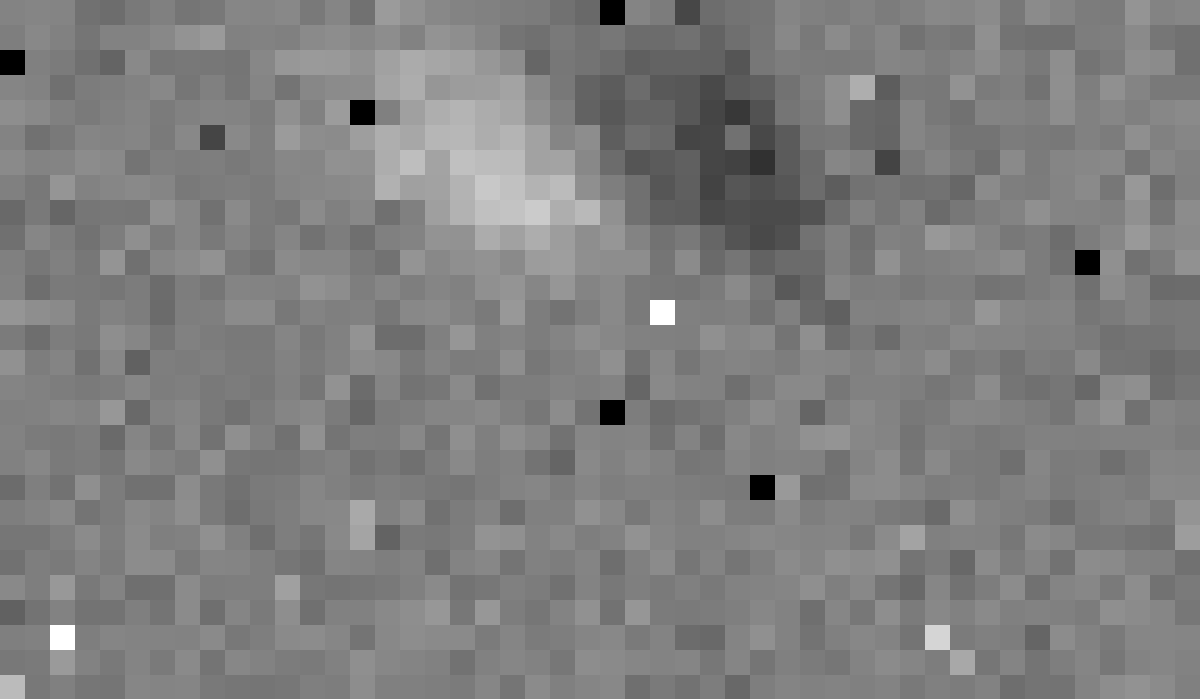}\\\vspace{0.07cm}
\includegraphics[width=0.23\textwidth]{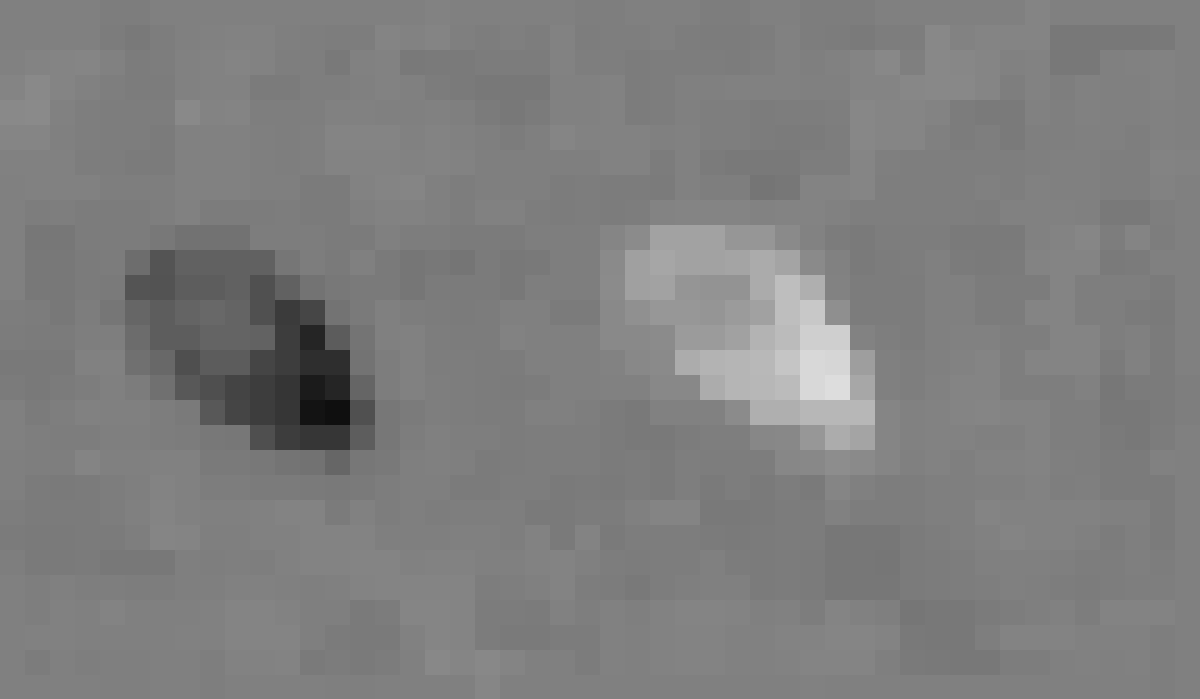}
\includegraphics[width=0.23\textwidth]{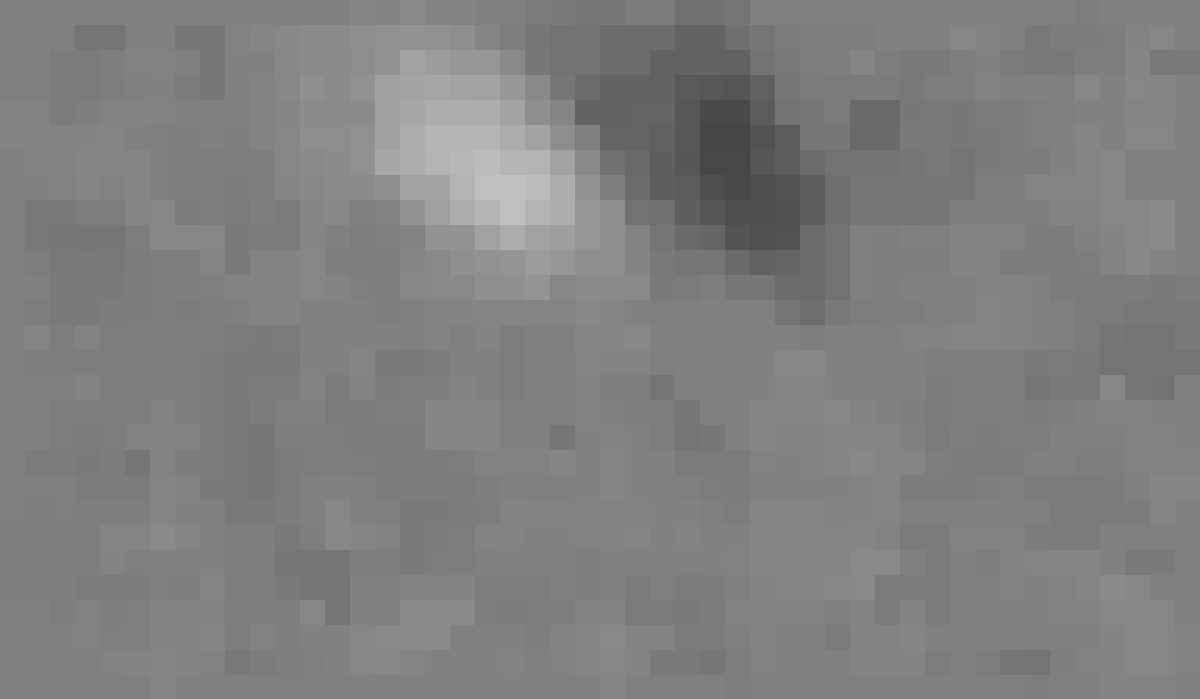}\\\vspace{0.07cm}
\includegraphics[width=0.23\textwidth]{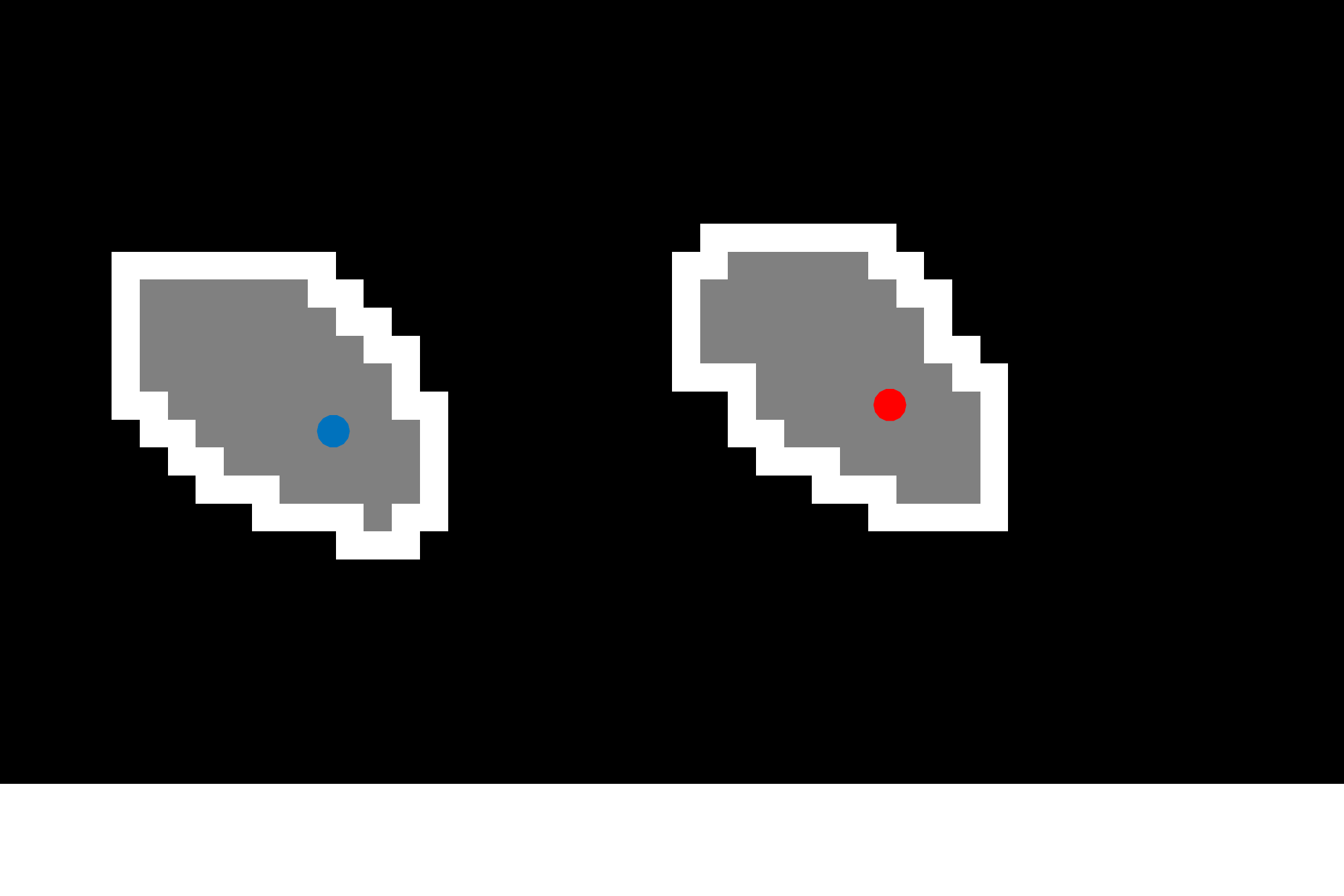}
\includegraphics[width=0.23\textwidth]{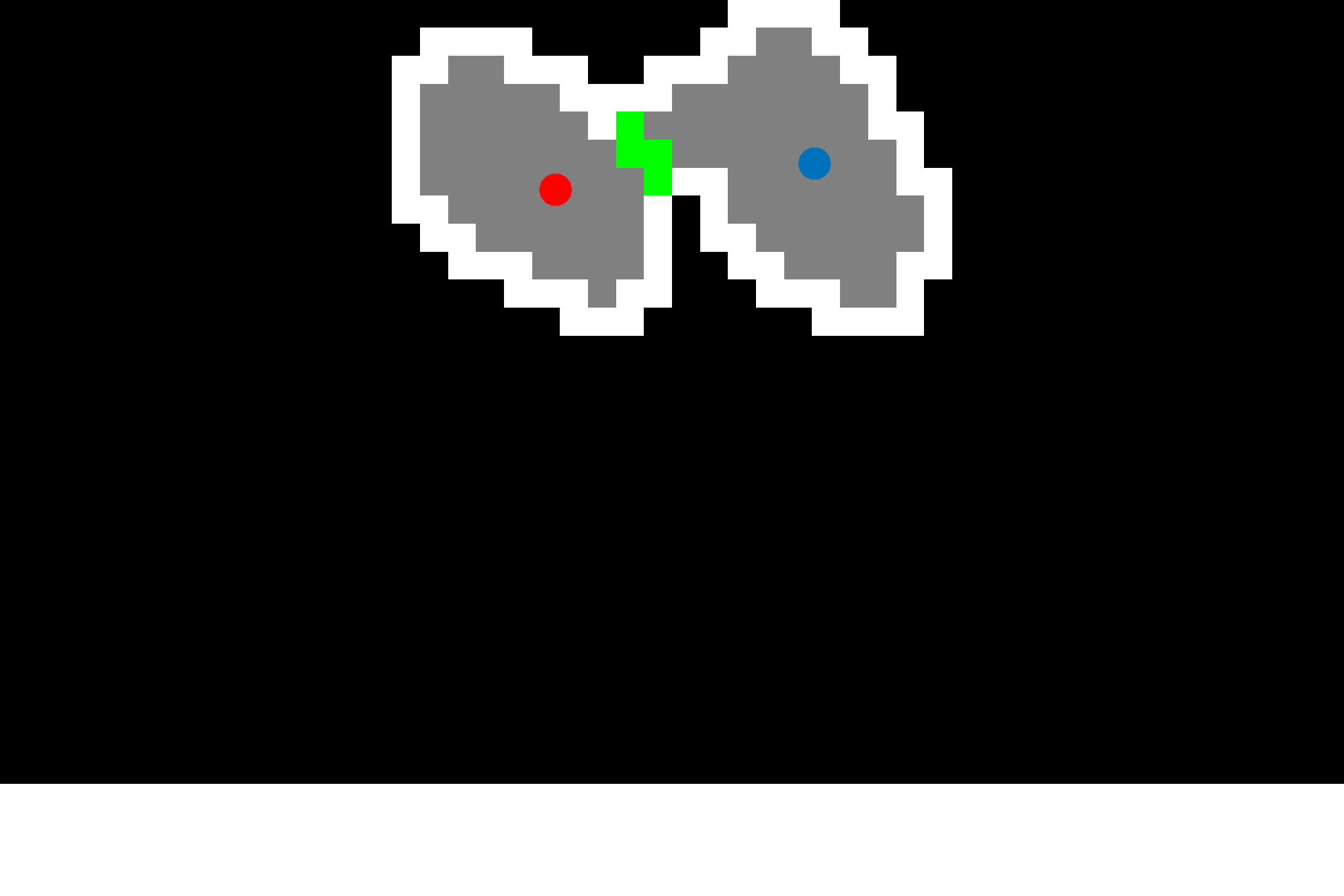}
\caption{Examples of images classified as useful (\textit{left}) and defective (\textit{right}) in the chopping algorithm for HD\,201078 = DT Cyg, BLb satellite, Cygnus I field, $T=$ 38$^\circ$C. \textit{From top to bottom}: $n$-th image; $(n-1)$-th image; difference image; median-filtered image; final stellar masks. The masks are shown in gray, dilation limit in white, overlapping pixels in green, centroid positions with red and blue dots for the positive and the negative profile, respectively.}
\label{chop_classify}
\end{figure}

The classification of images in chopping mode is based on the positions of centroids of the positive and negative stellar profiles in a difference image (Fig.\,\ref{chop_classify}). In particular, we use the difference between the coordinates of the two stellar centroids in the direction in which a satellite is moved in the chopping mode.\footnote{Since CTI smears the stellar profiles and hot pixels along the direction of charge transfer (i.e., along CCD columns) during the CCD readout, it was decided that the satellites working in the chopping mode would be moved roughly perpendicularly to this direction, along the $x$ coordinate.} Since the movements are done along the $x$ coordinate, the difference is denoted $\Delta_x$. A difference image is classified as useful if the following three conditions are fulfilled: (i) $\Delta_{\rm low} < \Delta_x < \Delta_{\rm high}$. The lower and upper limits for the difference in $x$ coordinate between the centroids of the positive and negative profiles are chosen to be $\Delta_{\rm low} = 0.25 X$ and $\Delta_{\rm high} = 0.75X$, respectively, where $X$ is the length of the subraster along the $x$ coordinate, (ii) the two stellar profiles do not overlap, and (iii) the positive profile does not touch the subraster edge\footnote{The negative profile can touch the raster edge because the photometry is performed using the positive profile.}. Otherwise, the image is classified as defective.

\begin{figure}[!t]
\centering
\includegraphics[width=0.48\textwidth]{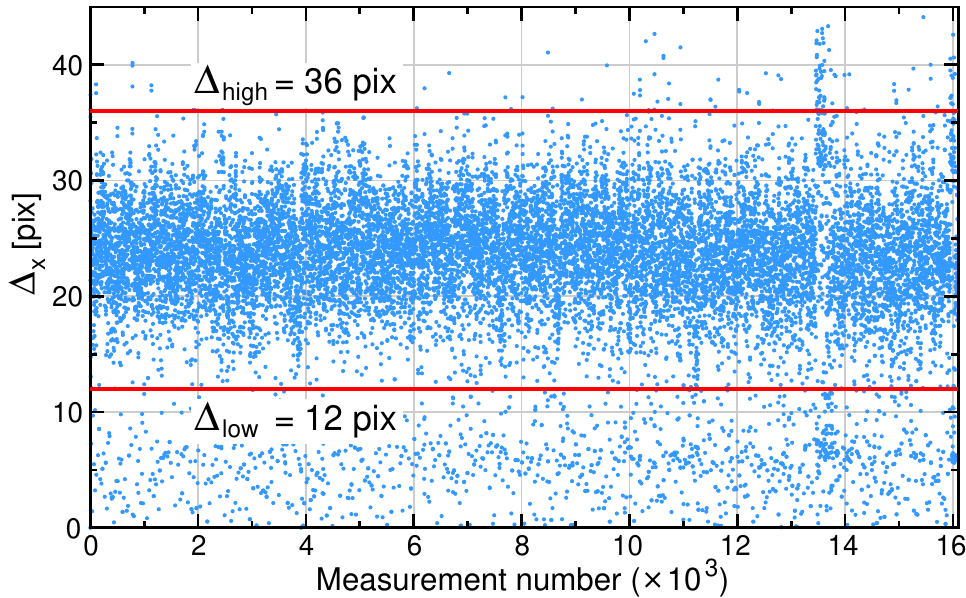}\\\vspace{0.2cm}
\includegraphics[width=0.15\textwidth]{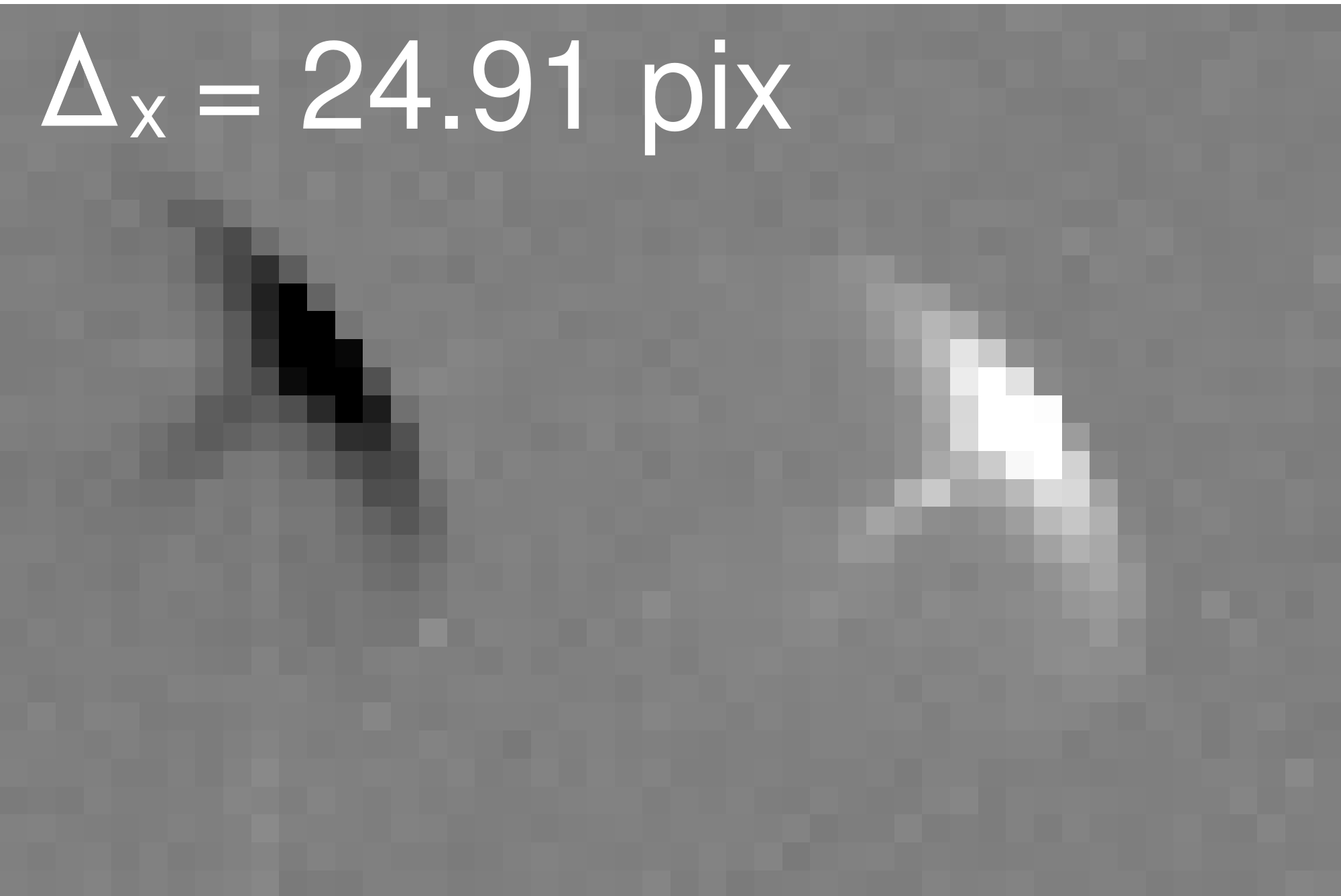}
\includegraphics[width=0.15\textwidth]{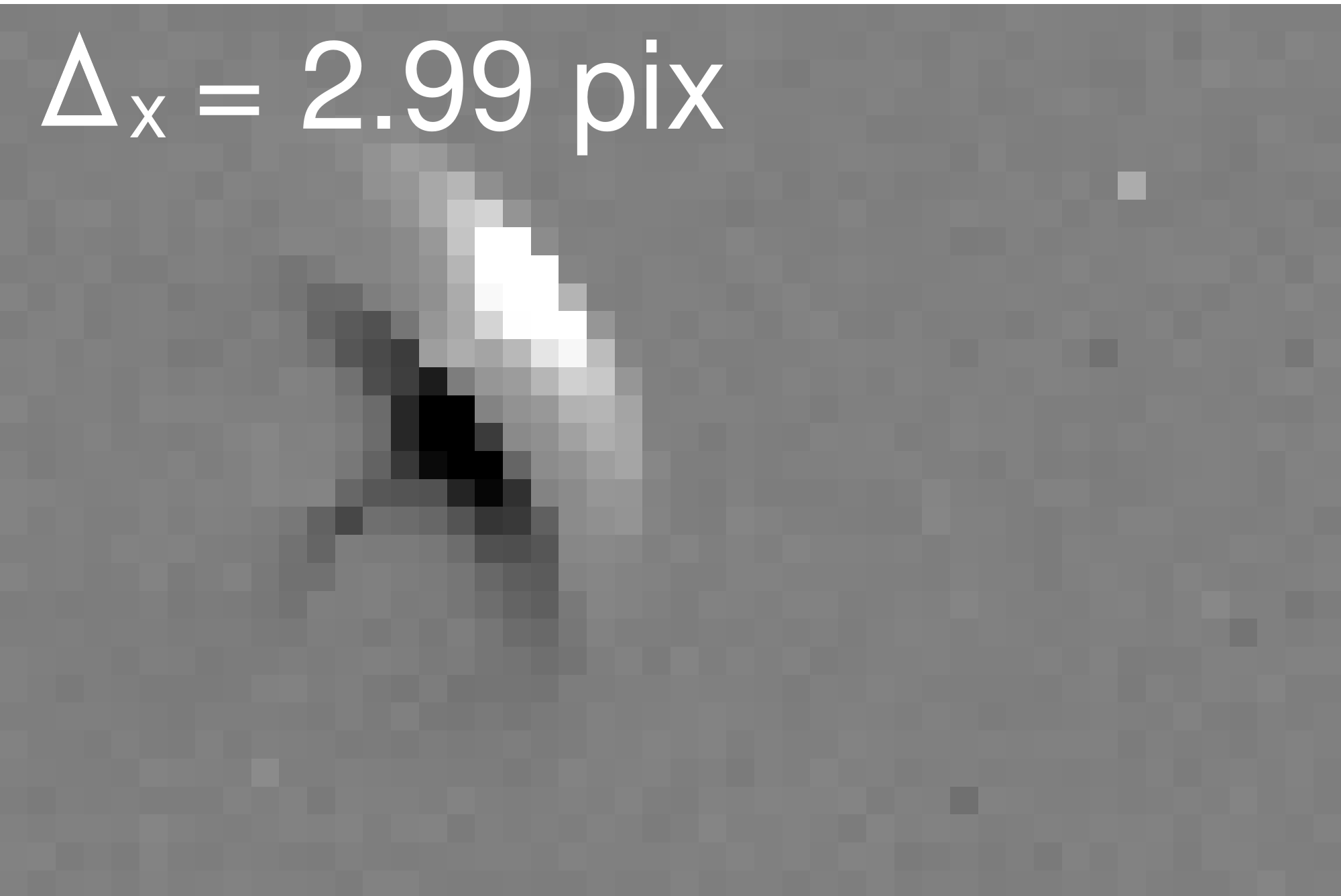}
\includegraphics[width=0.15\textwidth]{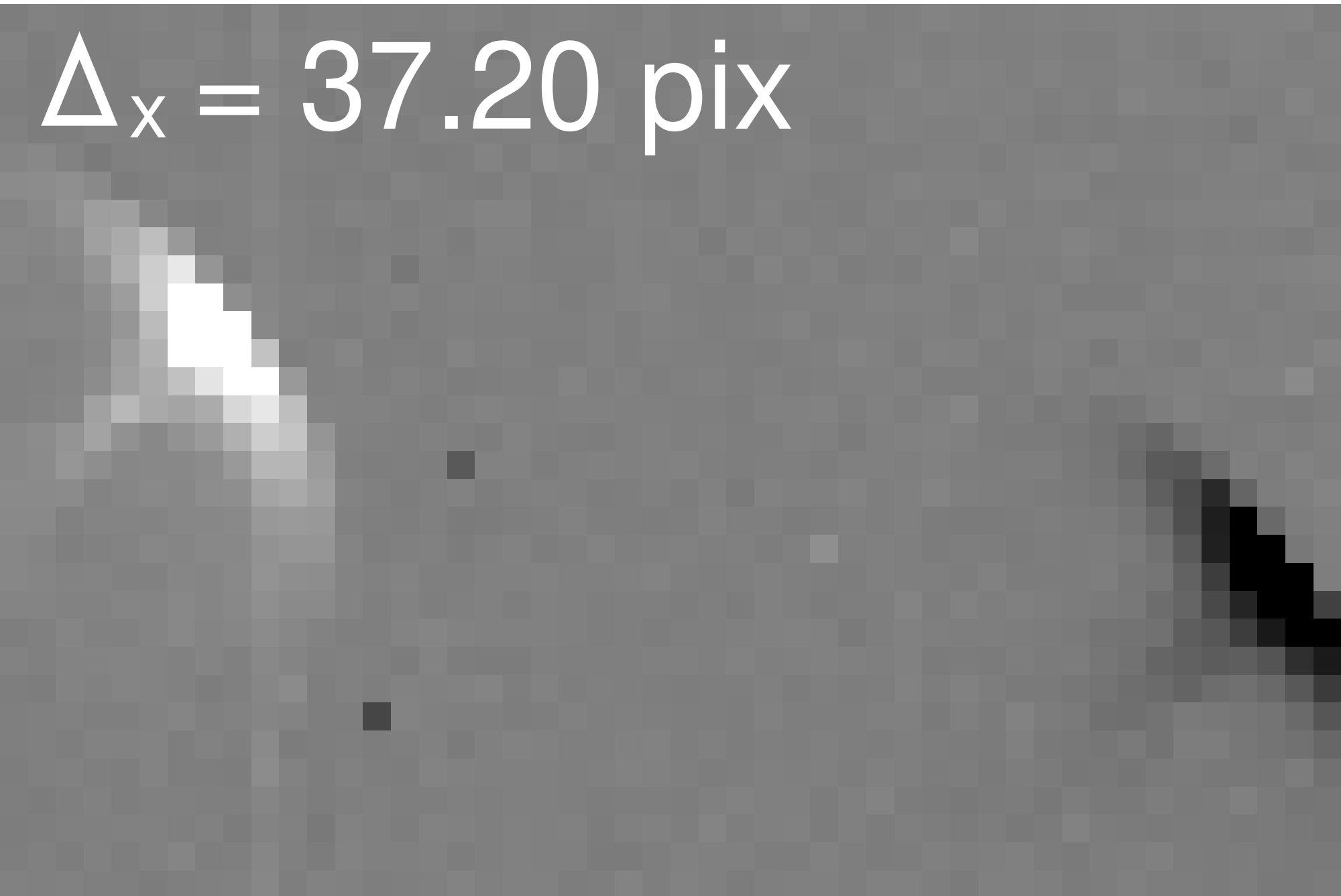}
\caption{Classification of images based on the value of $\Delta_x$. The top plot shows $\Delta_x$ for 19\,000 UBr measurements of HD\,138690 ($\gamma$~Lup) in the Sco I field. The two horizontal lines indicate the high ($\Delta_{\rm high}=$ 36\,pix $=$ 0.75$X$, $X=$ 48\,pix) and low ($\Delta_{\rm low}=$ 12\,pix $=$ 0.25$X$) cut-off thresholds. The three images shown below correspond to three cases, \textit{from left to right}: useful image ($\Delta_{\rm low}<\Delta_x =$ 24.91\,pix $<\Delta_{\rm high}$); image with overlapping profiles ($\Delta_x =$ 2.99\,pix $<\Delta_{\rm low}$); image with profiles too far from each other ($\Delta_x =$ 37.20\,pix $>\Delta_{\rm high}$).}
\label{chop_classificationplot}
\end{figure}

To obtain the positions of centroids, the difference image is first median-filtered with a 3\,$\times$\,3 pixel sliding window and then thresholded (above 100\,ADU for the positive profile and below $-$100\,ADU for the negative profile, respectively).  The purpose of filtering was to reduce occasional flickering of hot pixels. The thresholds are used to obtain the two stellar masks, positive, $M_+$, and negative, $M_-$, see Fig.\,\ref{chop_classify}. For both the positive and negative stellar profile, the centroid is calculated in the median-filtered and thresholded image. The differences, $\Delta_x$, for a sample set of images are presented in Fig.\,\ref{chop_classificationplot}. 

In order to check if the positive and negative profiles overlap, the two masks are independently dilated by one pixel in each direction. Then, if there are pixels that belong to both dilated masks, the image is classified as defective. The same is done if any pixel of an extended, positive mask touches the subraster edge. Figure \ref{chop_classify} shows two examples of the classification process for one of the faintest stars observed by BRITEs ($V=$ 5.8~mag) and one of the highest temperatures of the detector ($T=$ 38$^\circ$C) resulting in a strong dark-noise component. Such images would be lost in the stare pipeline because of the large number of hot pixels. Almost all hot pixels cancel out in the resulting difference image. Those which remain occur as a consequence of the RTS phenomenon. On the left-hand side, the process that leads to classifying an image as a useful one is shown. On the right-hand side, we obtain a defective image due to the proximity of the profiles. 

\subsection{Photometry with optimization}\label{phopt}
\begin{figure}[!t]
\centering
\includegraphics[width=0.23\textwidth]{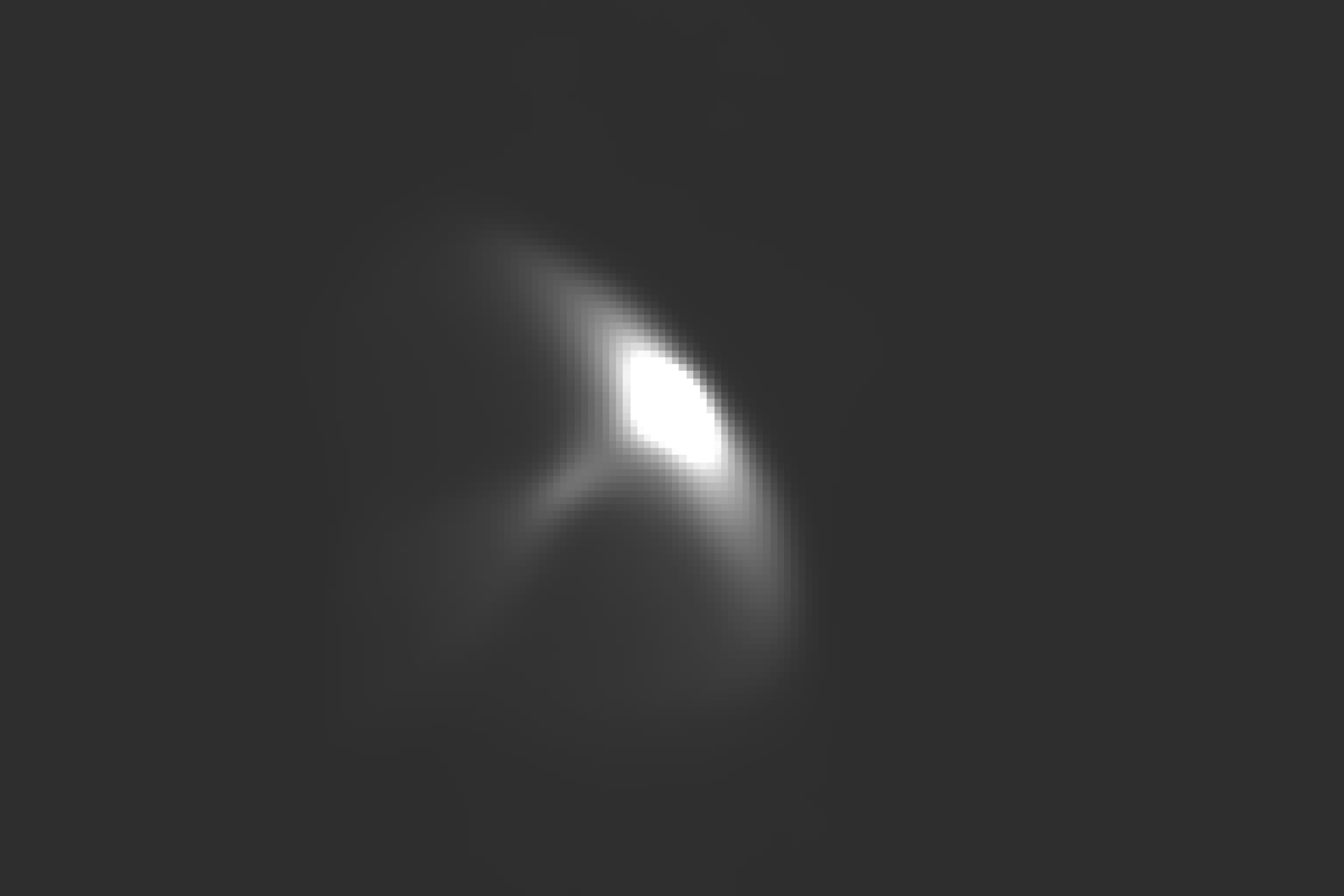}
\includegraphics[width=0.23\textwidth]{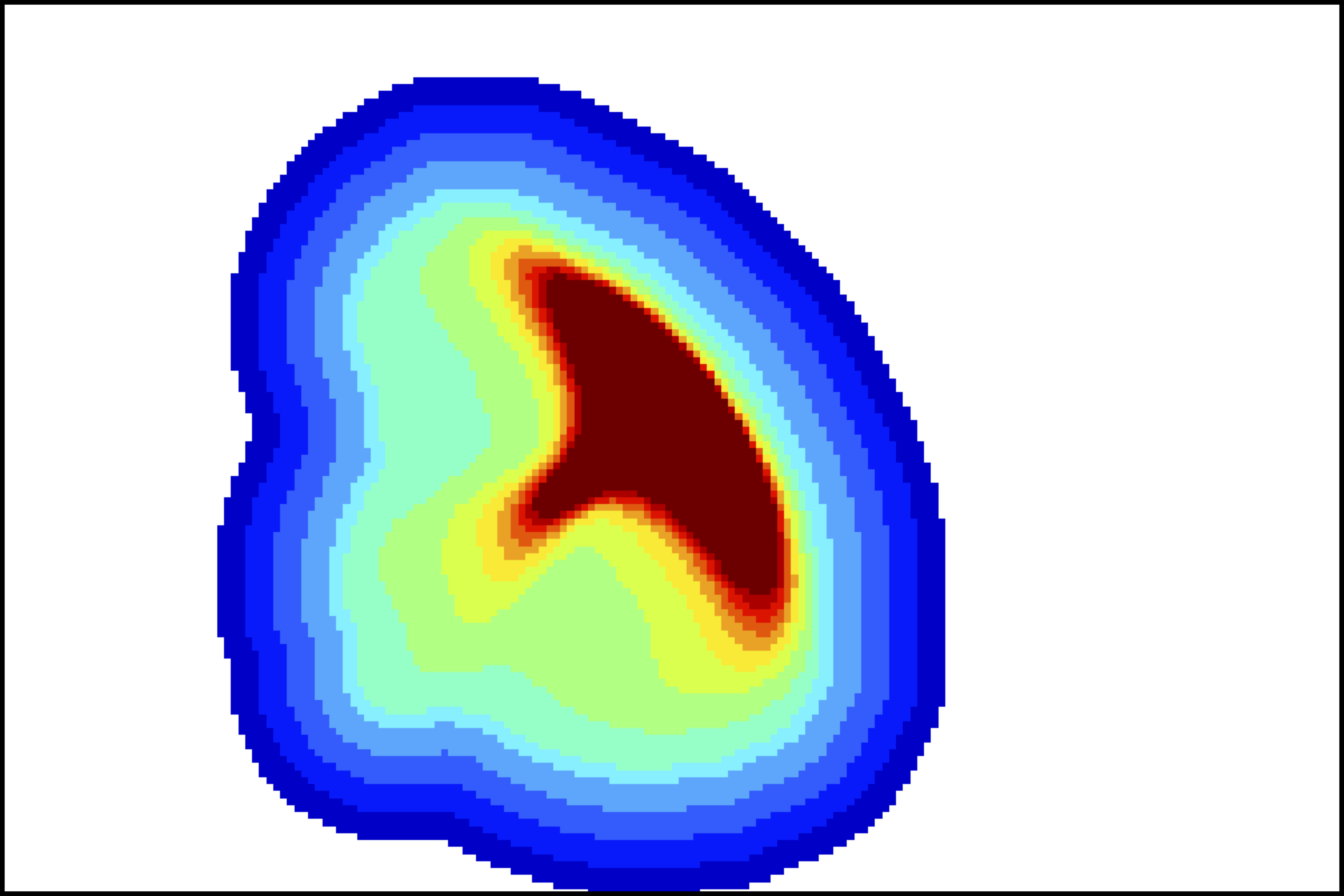}
\caption{\textit{Left:} Stellar profile template defined at a four times finer grid than the original images for the UBr observations of $\gamma$~Lup in the Sco I field. \textit{Right:} Thresholded apertures; each colour corresponds to a different aperture, the thresholds range from 25 to 200~ADU with a step of 25 ADU, with another threshold at 10~ADU.}
\label{chop_apertures}
\end{figure}

Having classified the images, a high-resolution stellar profile template is created from all useful images. For this purpose, for each difference image only the $M_+$ mask is retrieved and linearly interpolated onto a four times finer grid, i.e., each real pixel is replaced by a 4\,$\times$\,4 grid of square sub-pixels. Then, the profiles were co-aligned using the information on centroids and averaged. An example of the final stellar profile template is shown in Fig.\,\ref{chop_apertures}.

Photometry for observations in stare mode was made with circular apertures. Given the complex shapes of stellar profiles, such apertures may not be optimal from the point of view of the error budget, since circular apertures frequently include many pixels with very low or even no signal from a star. We therefore decided to obtain photometry in the pipeline for chopping data both with circular and non-circular apertures. The non-circular apertures were defined in the following way. In the first step, the apertures are created from the high-resolution template by simply thresholding it with levels ranging from 25 to 200\,ADU with a step equal to 25\,ADU plus another threshold at 10\,ADU. In this way, nine apertures are created. Furthermore, the largest aperture (obtained with a threshold equal to 10\,ADU) is enlarged by a dilation ranging from 2 to 18 sub-pixels with a step of 4 sub-pixels (the operation is performed on a finer grid). This gives five additional apertures. In effect, we define 14 non-circular apertures. An example of their shapes is shown in Fig.\,\ref{chop_apertures}. In addition, a series of 11 circular apertures with radii in the range between 4 and 44 sub-pixels with a step of 4 sub-pixels is used.
\begin{table*}[!t]
\centering
\caption{Pre-flight characteristics of BRITE CCD detectors.}
\begin{tabular}{lrrrrrr}
\hline\hline\noalign{\smallskip}
Temperature [$^\circ$C] &$-$20&0&10&20&30&60\\
\noalign{\smallskip}\hline\noalign{\smallskip}
Inverse gain, $G$ [e$^-$\,ADU$^{-1}$] &3.43&3.42&3.49&3.44&3.40&3.60\\
Dark current generation rate, $D_0$ [e$^-$\,s$^{-1}$\,pixel$^{-1}$] &0&2&8&21&49&534\\
Read-out noise, per pixel, $r_{\rm n}$ [e$^-$\,pixel$^{-1}$]&13&13&14&16&19&63\\
\noalign{\smallskip}\hline
\label{tab_CCD}
\end{tabular}
\end{table*}

Having defined the apertures, the final photometry is made with positive profiles in difference images. For each series of difference images, the photometry is done with all 25 apertures, both circular and non-circular. Each aperture is first shifted to the centroid of a positive profile in a difference image. Then, the aperture is binned to the lower, original resolution, and signals from pixels in a difference image confined by the aperture are co-added. Signals from pixels falling partially into aperture are added proportionally to the area of this pixel falling into the aperture. Ultimately, time-series photometry in all 25 apertures is obtained. Similarly to the pipeline for the stare mode of observing, the last step of the process consists of applying corrections for intra-pixel variability in the same way as described in Sect.\,\ref{intra}. Next, the optimal aperture is chosen by calculating the $Q_m$ parameter; see Eq.\,(\ref{Qm}). Photometry with the smallest $Q_m$ is regarded as optimal. We observed that circular apertures were chosen as optimal only in the rare cases when the stellar profile was round or when observations were strongly affected by tracking problems (stellar profile varied between exposures). 

The pipeline described in this section was applied to all BRITE chopping data which have been reduced up to now, i.e.~ten fields (see Table \ref{rel-flds}). The only exception were test data in the chopping mode obtained in the Perseus field which were reduced with the pipeline for the stare mode of observing (Sect.\,\ref{stare}). The resulting photometry is now available as Data Releases 3 (DR3), 4 (DR4), and 5 (DR5). A detailed explanation of data releases is available in Appendix \ref{drs}. The pipeline is essentially the same for these three data releases, but outputs include different parameters for decorrelation (Table \ref{datar}). The only change to the pipeline, applied to the data provided in DR5, is a compensation for a magnitude offset between the two positions in chopping during the optimization phase of reduction. This change could have affected the selection of optimal apertures and therefore the final photometry.

\section{The quality of BRITE photometry}\label{perf}
The final quality of the BRITE photometry must be assessed from real data. It is, however, good to know what can be expected given the parameters of detectors and observing strategy, and check if the real data meet expectations. It is also important to know if the quality of the photometry meets the mission requirements defined in Paper II. We therefore start with the theoretical expectations which are given in Sect.\,\ref{theo} followed by the estimation of the uncertainties from real data.

\subsection{Statistical uncertainties}\label{theo}
A standard formula for the photometric uncertainties associated with CCD photometry of stars usually includes four sources of noise: Poisson (or shot) noise related to the signal from a star ($\sigma_{\rm star}$), Poisson noise related to sky background ($\sigma_{\rm sky}$), dark current noise ($\sigma_{\rm dark}$), and read-out noise ($\sigma_{\rm RON}$). Assuming that all sources of noise are independent, the formula for the total theoretical noise, $\sigma_{\rm N}$, can be expressed as:
\begin{equation}
\sigma_{\rm N}^{2} = \sigma_{\rm star}^2 + \sigma_{\rm sky}^2 + \sigma_{\rm dark}^2 + \sigma_{\rm RON}^2. 
\label{eq_err}
\end{equation}
The only component in Eq.\,(\ref{eq_err}) that can be neglected for BRITE data is $\sigma_{\rm sky}$ since for such bright stars as those observed by BRITE, the background contribution to error budget is negligible. The total signal measured within the apertures for stars observed by BRITEs, $S$, ranges between 10$^4$ and 2\,$\times$\,10$^{6}$ electrons (e$^-$). This corresponds to $\sigma_{\rm star} = \sqrt{S} =$ 100\,--\,1500~e$^-$, depending on the stellar magnitude. 

In order to estimate the remaining two sources of noise, it is good to recall the temperature dependence of the detector-related parameters. They were taken from Tables 3, 4, and 5 of Paper II and are given in Table \ref{tab_CCD}. The temperature variations are relatively large for BRITE satellites. Taking into account all five satellites and 11 fields observed up to August 2016, the temperatures of the CCD detector, $T$, ranged between 2$^\circ$C and 41$^\circ$C. Using the data from Table \ref{tab_CCD}, we derived the following empirical relations for the dependencies of the dark current generation rate measured on the ground, $D_0$, and the read-out noise, $r_{\rm n}$, on temperature:
\begin{equation}
D_0(T) = 10^{0.0365T + 0.562},
\label{eq_d0}
\end{equation}
\begin{equation}
r_{\rm n}(T) = 12.924 + 0.20213T - 0.010175T^2 + 0.00034524T^3,
\label{eq_rn}
\end{equation}
where $T$ is given in $^\circ$C. Given the quoted range of temperatures, we get a range of $D_0$ between 4 and 110\,e$^{-}$\,s$^{-1}$ per pixel. Most of the BRITE data are taken with exposure time $t_{\rm exp} =$ 1\,s and the aperture radii, $R$, range between 2 and 12 pixels. The dark current contribution to Eq.\,(\ref{eq_err}) $\sigma_{\rm dark}^2 = AD_0t_{\rm exp}$, where $A$ is the aperture area in pixels. Given the values of $A$ and $D_0$, $\sigma_{\rm dark}$ ranges between 11 and 186 e$^-$. Finally, the contribution of the read-out-noise can be estimated from Eq.\,(\ref{eq_rn}). For the adopted ranges of temperature and aperture radius, $\sigma_{\rm RON} = \sqrt{Ar_{\rm n}^2}$ ranges between 69 and 485~e$^-$. 
\begin{table}[!t]
\centering
\caption{Some parameters of BRITE satellites related to error estimates. $D_0$ and $r_{\rm n}$ are given in the same units as in Table \ref{tab_CCD}.}
\begin{tabular}{lrrrrr}
\hline\hline\noalign{\smallskip}
Parameter &BAb & BLb & UBr & BTr & BHr \\
\noalign{\smallskip}\hline\noalign{\smallskip}
Median $T$ [$^\circ$C] & 23.1 & 20.6 & 26.3 & 15.0 & 15.9\\
Median $D_0$ & 25.4 & 20.6 & 33.3 & 12.9 & 13.9\\
Median $r_{\rm n}$ & 16.4 & 15.8 & 17.5 & 14.8 & 15.0 \\
Mean $A$ [pixels] & 134 & 190 & 98 & 125 & 119 \\
\noalign{\smallskip}\hline
\label{tab_sat}
\end{tabular}
\end{table}

Taking $t_{\rm exp}=$ 1\,s for most BRITE observations, the final formula for theoretical uncertainty expressed in e$^-$ is the following:
\begin{equation}
\sigma_{\rm N} = \sqrt{{S+A(D_0+r_{\rm n}^2)}}
\label{noise-th}
\end{equation}
or, if given in magnitudes:
\begin{equation}
\sigma_{\rm N}^{\rm mag} \approx 2.5\log\left(1 + \frac{\sqrt{{S+A(D_0+r_{\rm n}^2)}}}{S}\right).
\label{noise-th-mag}
\end{equation}

\begin{figure}[!t]
\centering
\includegraphics[width=0.48\textwidth]{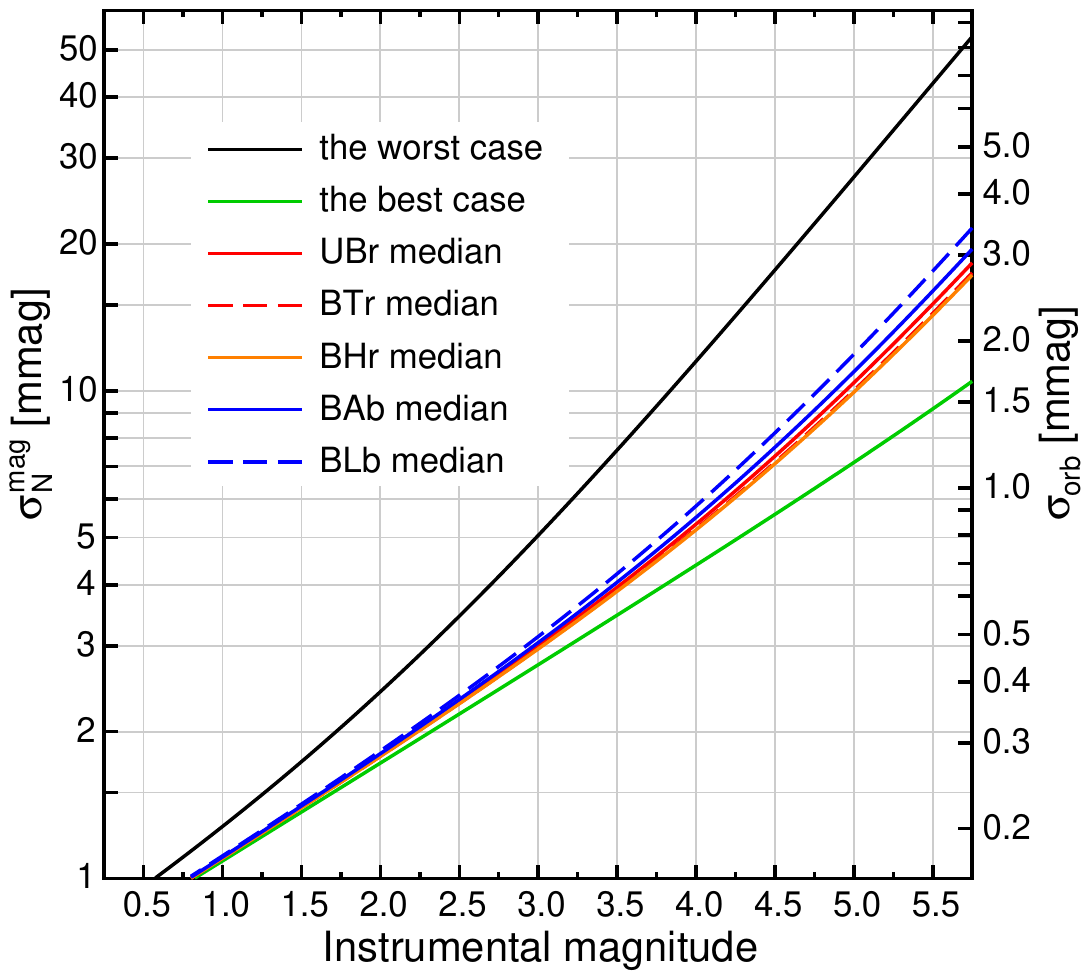}
\caption{Theoretical uncertainties for BRITE photometry for observations in stare mode, $\sigma_{\rm N}^{\rm mag}$, assuming the smallest (green) and the largest (black) contribution to $\sigma_{\rm dark}$ and $\sigma_{\rm RON}$ in Eq.\,(\ref{noise-th-mag}). The dependencies for all five satellites are also shown assuming the median parameters given in Table \ref{tab_sat}. The $\sigma_{\rm orb}$ shown on the right-hand side ordinate equals to $\sigma_{\rm N}^{\rm mag}/\sqrt{\mbox{40}}$.}
\label{err_budget}
\end{figure}

The quoted numbers show that for the range of temperatures under consideration, $\sigma_{\rm RON} > \sigma_{\rm dark}$. Theoretically, read-out noise is therefore the dominating source of noise for faint stars (in the BRITE magnitude range) while for bright stars it is the shot noise that principally determines the error budget. Figure \ref{err_budget} shows the theoretical uncertainty $\sigma_{\rm N}^{\rm mag}$ as a function of instrumental magnitude defined as $m_{\rm inst} = -\mbox{2.5}\log(S[\mbox{e}^-]) + \mbox{16.0} \approx -\mbox{2.5}\log(S[\mbox{ADU}]) + \mbox{14.6}$. The constant 16.0 was chosen to tie very roughly the BRITE red filter instrumental magnitudes to $r$ magnitudes of the SDSS $ugriz$ system using the calibration of \cite{2008PASP..120.1128O}. The choice was made due to the similarity of the BRITE red filter and SDSS $r$ passbands. The curves for different satellites were calculated for median values of $D_0$ and $r_{\rm n}$, which are temperature dependent and therefore vary from one satellite to another (see Paper II). They are given in Table \ref{tab_sat} together with mean values of $A$. All parameters in this table were derived for the observations in stare mode. While $\sigma_{\rm N}^{\rm mag}$ represents the theoretical uncertainty of a single measurement obtained with a 1-second BRITE exposure, the right-hand ordinate in Fig.\,\ref{err_budget} shows the uncertainty for the average of all measurements within an orbit. The number of measurements per orbit ranges between several and over 160 and depends on the satellite and field\footnote{Sometimes a satellite switches between two fields in a single orbit. This usually results in a smaller number of points per orbit in each of the two fields.}, but a mean value amounts to about 40 measurements per orbit. Therefore, in Fig.\,\ref{err_budget}, $\sigma_{\rm orb} = \sigma_{\rm N}^{\rm mag}/\sqrt{\mbox{40}}$. As one can see, the $\sigma_{\rm orb} <$ 1~mmag theoretical uncertainty can be achieved for stars brighter than 4th magnitude.

Equations (\ref{noise-th}) and (\ref{noise-th-mag}) are valid for the stare mode of observing. In the chopping mode, the photometry is done with difference images, so that the dark noise and read-out noise contributions in Eq.\,(\ref{noise-th}) have to be multiplied by a factor of $\sqrt{2}$ and the equation for the chopping mode has the following form:
\begin{equation}
\sigma_{\rm N}^{\rm mag} \approx 2.5\log\left(1 + \frac{\sqrt{{S+2A(D_0+r_{\rm n}^2)}}}{S}\right).
\label{noise-th-mag-chop}
\end{equation}

\begin{figure}[!t]
\centering
\includegraphics[width=0.49\textwidth]{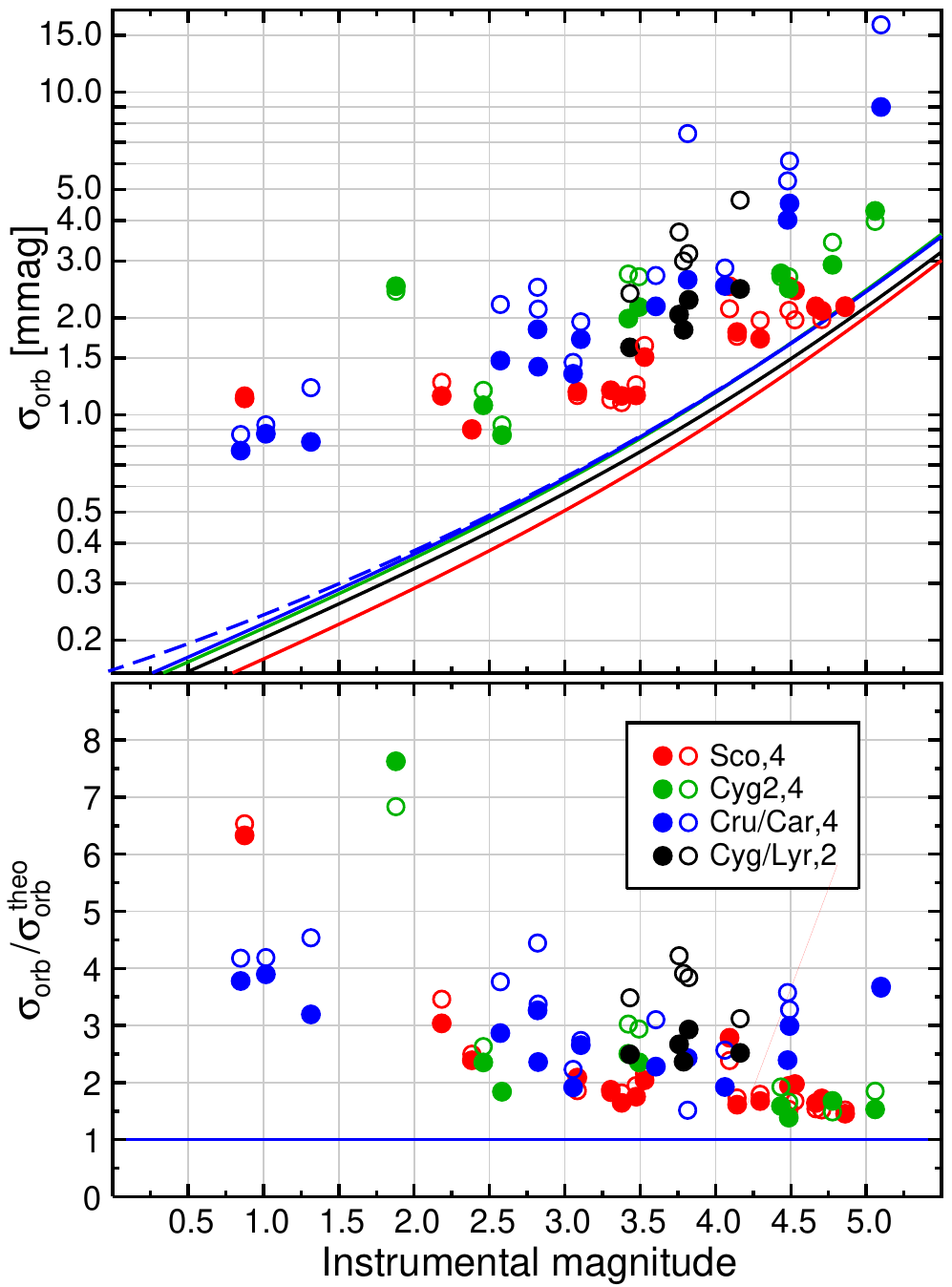}
\caption{{\it Top:} Median standard deviation per orbit, $\sigma_{\rm orb}$, estimated from test UBr data (see Sect.\,\ref{normvschop}) for stars in the Sco, Cyg~2, Cru/Car, and Cyg/Lyr fields, obtained in stare (open circles) and chopping mode (filled circles). Colours used for stars in the four fields are explained in the inset. The continuous lines are theoretical relations for average CCD temperature, average aperture, and average number of points per orbit, calculated from Eq.\,(\ref{noise-th-mag-chop}). The colours are the same as used for plotting symbols. The dashed line is the theoretical line with the flat-field noise of 0.4~mmag per point added. This line is shown only for the Cru/Car field. {\it Bottom:} The ratio of $\sigma_{\rm orb}$ and $\sigma_{\rm orb}^{\rm theo}$, the theoretical values of standard deviation per orbit, calculated for individual sizes of optimal aperture. Symbols are the same as in the top panel.}
\label{sig-mag-ubr}
\end{figure}

Equation (\ref{eq_err}) does not include some other possible sources of noice. One of them could be the contribution related to the lack of flat-fielding \citep{2009MNRAS.397.2049S}, which cannot be done properly in orbit. Using the pre-flight flat-fields obtained in the UTIAS-SFL\footnote{University of Toronto Institute for Aerospace Studies, Space Flight Laboratory.} clean room by a fully integrated instrument aboard UBr, we estimated that the variation of sensitivity over the CCD chip is constant to within $\pm$0.27\% rms. There is no known mechanism, which would increase this value in orbit. We assumed that the stellar profile is a truncated Gaussian with a 3-$\sigma$ radius and covers 100 pixels, a typical value of stellar area in BRITE photometry (Fig.\,\ref{classification1}). The profile was moved randomly on a pixel grid. We obtained that the noise component due to the lack of flat-fielding can be estimated to approximately 0.4\,mmag for a single measurement and about 0.07\,mmag for orbit-averaged measurements if we adopt that 40 points per orbit are measured. This is well below the Poisson noise, even for the brightest objects. A similar test was performed with several high-resolution profile templates obtained from the real BRITE images. The result was similar; the noise for orbit-averaged measurements did not exceed 0.1~mmag. More importantly, the utilized decorrelations with centroid positions effectively compensate for most of the possible effects due to the lack of flat-fielding, especially in the chopping mode. The final comparison of scatter in Sect.\,\ref{actper} is performed with decorrelated data. Thus, the 0.1\,mmag per orbit flat-field noise should be regarded only as an upper limit. Consequently, this source of noise was recognized as negligible and not taken into account in Eq.\,(\ref{eq_err}); see also Fig.\,\ref{sig-mag-ubr}. 

We also realize that for BRITE observations a significant source of noise is impulsive noise. Nevertheless, we do not include it into the noise equation either because it is of discrete nature (hot pixels), strongly depends on the temperature and time (number of CCD defects increases with time), and the associated RTS phenomenon has a complicated character. Therefore, it is very hard to characterize this source of noise with real numbers. We regard this source of noise as the main factor causing the real scatter to be higher than expected (Sect.\,\ref{actper}).

\subsection{Stare vs.~chopping mode of observing: the assessment of quality}\label{normvschop}
A simple comparison of Eq.\,(\ref{noise-th-mag}) and (\ref{noise-th-mag-chop}) shows that the stare mode of observing should lead to a smaller noise level and thus better photometry than the chopping mode. As we will show below, usually the opposite is true. This can be explained by the presence of an additional source of noise, not accounted for by Eq.\,(\ref{eq_err}). This is an impulsive noise which occurs in the form of hot pixels and depends primarily on temperature: higher temperatures lead to both larger density and higher amplitude of the impulsive noise. In addition, as explained in Sect.\,6 of Paper II, the impulsive noise increases with time because of the growing number of CCD defects caused by proton hits. A detailed study of the temporal behaviour of the impulsive noise is out of the scope of this paper. However, in order to justify the decision of switching from the stare to the chopping mode of observing, we show here how photometry obtained for the same data with the two pipelines compares. For this purpose, we ran both pipelines using data obtained in the chopping mode. This is because the pipeline developed to obtain photometry from images made in the stare mode can be applied to the chopping data, while the opposite is not possible. For simplicity, from now on we will use the notions `stare pipeline' and `chopping pipeline' for the two pipelines presented in this paper. In order to make the comparison, we chose six setups\footnote{Data from a given satellite pointing are split into parts, called setups, during reductions. See Appendix \ref{drs} for the explanation of setups.} of the UBr satellite runs in the Scorpius I, Cygnus II, Crux/Carina I, and Cygnus/Lyra I fields. The total span of the data is 543~d, between March 20, 2015 and September 13, 2016. For these data, the temperature of the CCD detector ranged between 14.5 and 36.1\,$^\circ$C. The observations were performed with a constant exposure time of 1\,s. The data for 16 stars in the Sco I field\footnote{We have excluded five stars from these tests, one because it is a close visual double, the remaining because of insufficient data points.}, 9 stars in the Cyg II field, 14 stars in the Cru/Car I field, and 5 stars in the Cyg/Lyr I field were reduced independently with the two pipelines described in Sect.\,\ref{stare} and \ref{chop}.
\begin{figure*}[!t]
\centering
\includegraphics[width=\textwidth]{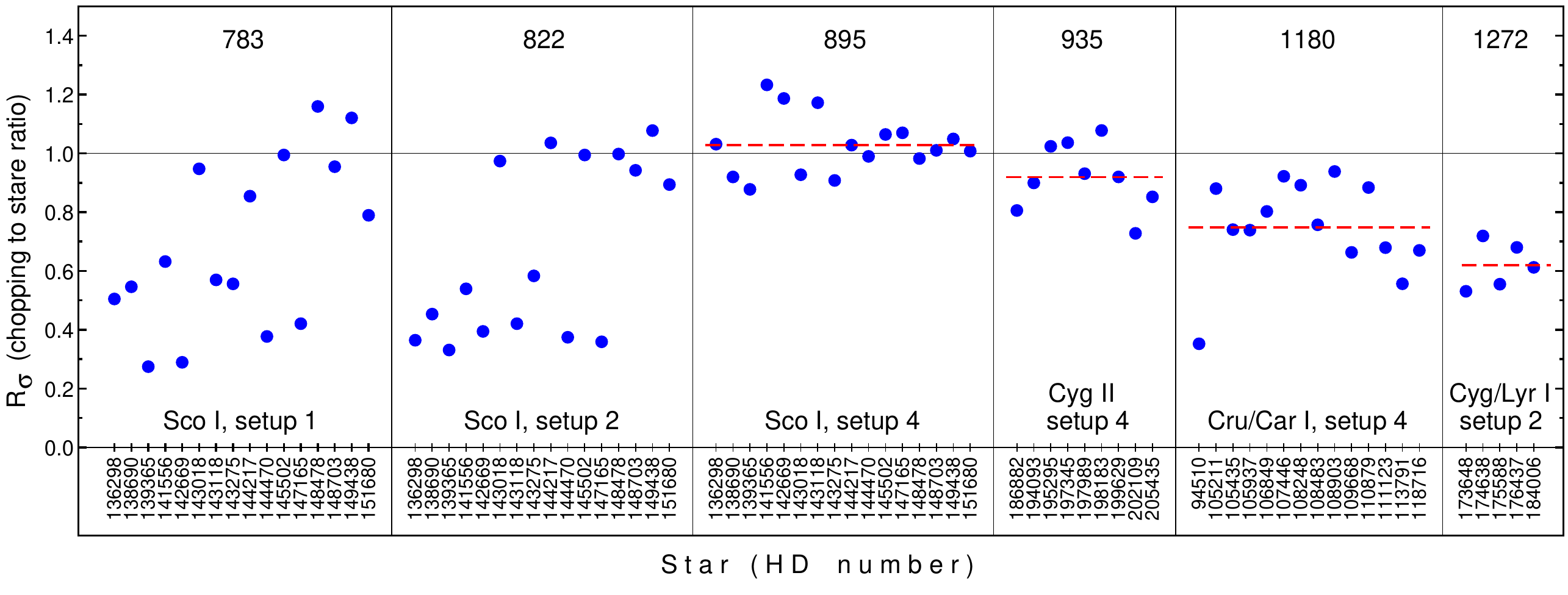}
\caption{The ratio of standard deviations, $R_\sigma$ (blue dots) defined as explained in the text, for UBr observations of 16 stars in the Scorpius I field (setups 1, 2, and 4), 9 stars in the Cygnus II (setup 4) field, 14 stars in the Crux/Carina I field (setup 4), and 5 stars in the Cygnus/Lyra I field (setup 2). The number at the top of each panel is the number of days elapsed since the launch of UBr, for the mid-time of a given setup. The horizontal red dashed lines denote the mean values of $R_\sigma$ for the last four test setups.}
\label{chop-norm}
\end{figure*}

The quality of the photometry is best characterized by its internal scatter which we evaluate using orbit samples. The reason for this is that the scatter in the raw photometry depends on four factors: (i) the sources of noise summarized by Eq.\,(\ref{eq_err}) and described in Sect.\,\ref{theo}, (ii) impulsive noise, not included in Eq.\,(\ref{eq_err}), (iii) instrumental effects which occur as correlations of the measured flux with different parameters, see Appendix \ref{decor}, (iv) intrinsic variability. In order to focus the comparison of the two pipelines on the influence of the impulsive noise, the raw photometry needs to be corrected for factors (iii) and (iv) first. 

The reasons for the occurrence of correlations and the procedures of decorrelation have been described e.g.~by \cite{2016A&A...588A..55P} and \cite{2017arXiv170400576B}, but for completeness, we summarize the parameters used in decorrelations in Appendix \ref{decor}. The number of parameters for decorrelation has evolved with the subsequent data releases of BRITE photometry. As we mentioned earlier, presently there are four different data releases available. The format of the output files as well as the explanation of the parameters which are used for decorrelation is briefly described in Appendices \ref{drs} and \ref{decor}, respectively. Each data release provides a different number of parameters for decorrelation.

The test data mentioned above, obtained with the two pipelines, were independently corrected for factors (iii) and (iv) leaving a slightly different number of data points. The correction for the intrinsic variability, i.e.~factor (iv), was made in two steps. Using Fourier frequency spectra, all periodic signals were identified, fitted by least-squares and subsequently subtracted from the data. Then, all residual non-periodic signals (if present) were removed by means of detrending. Detrending was based on the interpolation between averages calculated in time intervals of 1~d with Akima interpolation \citep{1991ACMT...17..341A,1991ACMT...17..367A}. In order to make the final photometry directly comparable, we subsequently used only those data points which had times common to the two resulting light curves. The examined parameter was the mean scatter (standard deviation) per orbit. Figure \ref{chop-norm} shows the ratio of median standard deviations, $R_\sigma=\mbox{me\-dian}\{\sigma_{{\rm chop},i}\} / \mbox{median}\{\sigma_{{\rm stare},i}\}$, where $\sigma_{{\rm chop},i}$ and $\sigma_{{\rm stare},i}$ are standard deviations in the $i$-th orbit for data obtained with the stare and chopping pipelines, respectively. As can be seen from the figure, for the first two setups (Sco I, setup 1 and 2) there is a group of nine stars -- the same for both setups -- that have $R_\sigma=$ 0.2\,--\,0.7. For these stars, the chopping pipeline provides much better photometry than the stare pipeline. These stars are located in the areas affected by CTI; see Sect.\,6 of Paper II. In order to mitigate the problem, starting from mid-July 2015 all BRITE satellites were set to slower read-out, which effectively solved the CTI problem (Paper II)\footnote{It seems that switching to lower read-out solved only one problem related to the CTI, i.e.~smearing of images. We have recently discovered that signal for stars located in regions of strong CTI is partly lost due to the traps in the detector. The problem needs to be investigated in detail. At the time of writing it is not known yet how severely BRITE data are affected.}. 

The remaining four setups of UBr observations were obtained with the slower read-out. For the Sco I setup 4 observations, $R_\sigma=$~1.03\,$\pm$\,0.03, i.e.~is close to unity showing no advantage of chopping pipeline.  For the next three setups the mean $R_\sigma$ gradually decreases to 0.92\,$\pm$\,0.04 for Cyg II, 0.75\,$\pm$\,0.04 for Cru/Car I, and 0.62\,$\pm$\,0.04 for Cyg/Lyr I, see the dashed lines in Fig.~\ref{chop-norm}. This clearly shows that the longer time elapsed since the launch (the setups are presented chronologically), the better the results are obtained with the chopping pipeline in comparison with the stare pipeline. The decrease of $R_\sigma$ with time is related to the increasing number of hot pixels in BRITE detectors. Switching all BRITE satellites to the chopping mode of observing is therefore fully justified and may even allow to extend the mission beyond the point where standard photometry would not be possible.

\begin{figure}[!t]
\centering
\includegraphics[width=0.49\textwidth]{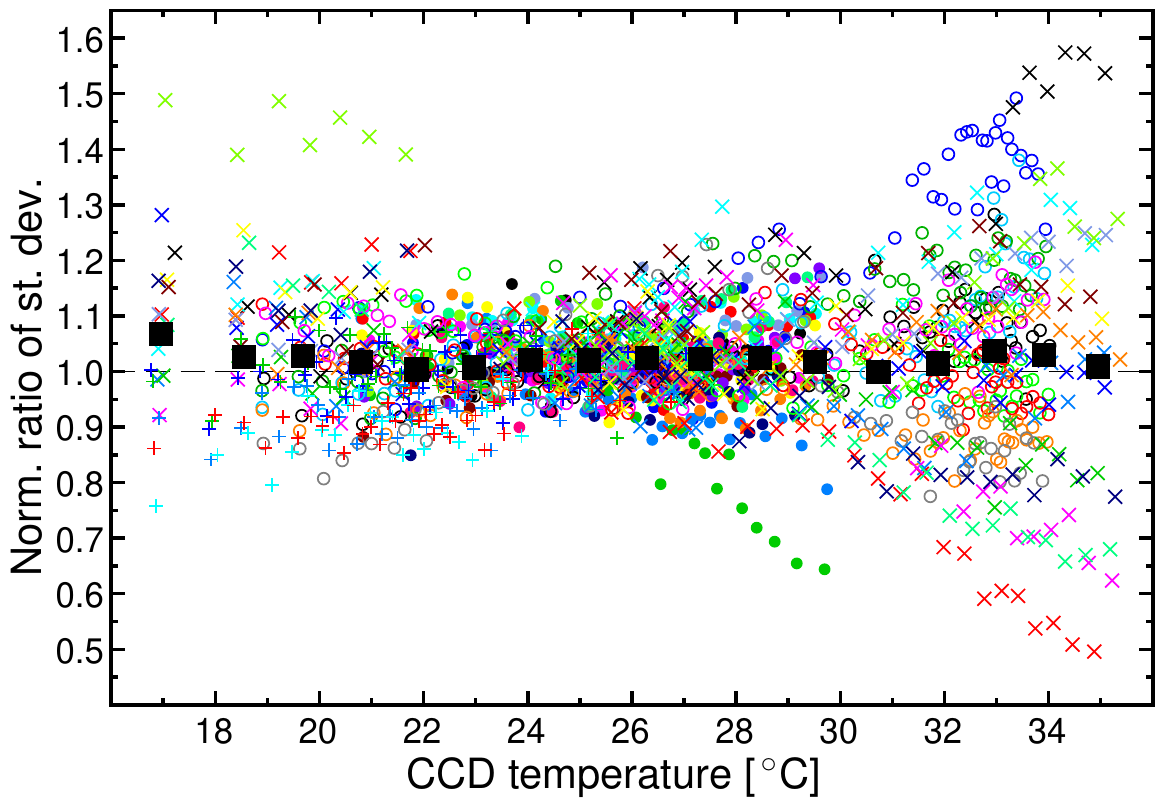}
\caption{Normalized (to CCD temperature of 25$^\circ$C) ratios of residual standard deviations (chopping to stare) for UBr data of the same stars as shown in Fig.\,\ref{chop-norm}. Different symbols for stars from different fields are used: dots for the Sco field, open circles for the Cyg~2 field, crosses for the Cru/Car field, and pluses for the Cyg/Lyr field. Large black squares are averages calculated from all points in 17 temperature intervals.}
\label{chop-norm-rat-t}
\end{figure}

We also expected that the advantage of using the chopping pipeline will be better enhanced at higher CCD temperatures because of the larger number of hot pixels. To make an appropriate comparison, we used the same data as in the previous test, but confined to the four setups not affected by CTI. For each star, the standard deviation as a function of CCD temperature was evaluated and then normalized to the scatter at 25$^\circ$C (using data with $T=$~25 $\pm$ 1$^\circ$C). The ratios of the resulting normalized standard deviations are shown in Fig.\,\ref{chop-norm-rat-t}. Surprisingly, the mean values of these ratios calculated from the dependencies for all 44 stars show practically no dependence on temperature (black squares in Fig.\,\ref{chop-norm-rat-t}). There are, however, strong differences between different stars. The ratio of standard deviations can either drop with temperature as for HD\,142669 = $\rho$~Sco (green dots) or HD\,105937 = $\rho$~Cen (red crosses) or increase as for HD\,194093 = $\gamma$~Cyg (blue circles). 

This shows that the benefits of the chopping pipeline compared to the stare pipeline do not neccessarily increase for higher temperatures. This can be explained as follows. First, the frequency of RTS increases with higher temperature; see Paper II and \cite{2016SPIE.9904E..1RP} for more details. Therefore, the probability that a hot pixel shows the same intensity in two successive images decreases. This introduces unwanted impulsive noise appearing in difference images, which is not corrected by the current chopping pipeline. Second, the dark current noise and readout noise are larger by a factor of $\sqrt{2}$ in difference images, cf.~Eqs.~(\ref{noise-th-mag}) and (\ref{noise-th-mag-chop}). This difference may also become more prominent at higher temperatures.

\subsection{Comparison between theoretical expectations and real performance}\label{actper}
It is also interesting to see how the real scatter in BRITE photometry compares with the calculations presented in Sect.\,\ref{theo}. A comparison has been done for the same test data for 44 stars as those presented in Sect.\,\ref{normvschop}. The results are shown in Fig.\,\ref{sig-mag-ubr}. One can see that for both observation modes the scatter of the photometry is larger than predicted by theory. For most stars the ratio $\sigma_{\rm orb}/\sigma_{\rm orb}^{\rm theo}$ ranges between 1 and 4 with a small increase towards brighter stars. Stars brighter than $\sim$1.5~mag are affected by known nonlinearities (see Sect.\,3.2.3 in Paper II), which combined with variable blurring and intra-pixel sensitivity variations results in a noise floor of approximately 0.7\,--\,1.0 mmag for the brightest objects. Therefore, the scatter in the photometry of such stars is much larger than predicted.

The main source of the degradation of BRITE photometry in comparison with theoretical expectations is the presence of high impulsive noise which has a non-stationary character. This is why $\sigma_{\rm orb}/\sigma_{\rm orb}^{\rm theo}$ values (lower panel of Fig.\,\ref{sig-mag-ubr}) are on average higher for stars in Cru/Car field in comparison with Sco field observed about a year earlier. The other important factor that contributes to the increase of the real scatter is a non-perfect stability of the satellites resulting in image blurring. Decorrelations applied to the data cannot fully account for these effects. With consecutive data releases (Appendix \ref{drs}) more parameters that can be used for decorrelations are provided. This means that we can believe that the later the data release the more effectively instrumental effects can be removed from the photometry. 

Below, we summarize in detail the possible factors that can lead to the degradation of the photometry in both observing modes. The items are marked with `[S]' or/and `[Ch]' indicating whether a given item is applicable to stare or chopping mode of observing, respectively.
\begin{itemize}
\item[$\bullet$][S] Interpolations introduce uncertainties in the estimation of the intensity of the replaced pixels. Their quality is decreased by the changes of the temperature of CCD, which induce subtle variations of stellar profiles.
\item[$\bullet$][S] An optimized selection of the hot pixels in master dark frames is always a trade-off. Too many pixels classified as hot would decrease the photometric quality, but leaving too many hot pixels would increase RTS-related noise.
\item[$\bullet$][S \& Ch] The precision of the centroid position, based on the centre of gravity, suffers from dark current noise and interpolation errors.
\item[$\bullet$][S] The shapes of stellar profiles are usually highly non-circular (see Fig.\,\ref{PSFs}) and change due to the changes of the temperature of the optics. The optimal circular aperture does not always cover the whole stellar profile. In consequence, small differences in the position of the centroid can lead to a non-negligible change of the flux measured in the aperture.
\item[$\bullet$][S \& Ch] To some extent, all images are blurred. Since we use a constant aperture, a different part of the total flux is measured, depending on the amount of blurring. This effect is largely compensated by decorrelation with blurring parameters, but some excess of scatter due to this effect can be expected.
\item[$\bullet$][Ch] The background may change between consecutive difference images, especially when scattered light from either the Moon or Earth changes rapidly in time. Starting from Data Release 5 this effect can be compensated for by a decorrelation with respect to parameter APER0; see Appendix \ref{decor}.
\item[$\bullet$][Ch] The RTS-related noise increases with temperature because transitions between many meta-stable states happen more frequently, reducing the effectiveness of impulsive noise cancellation in difference images.
\end{itemize}

\section{Conclusions and future work}\label{concl}
The main goal of BRITE-Constellation is to provide precise photometry of bright stars. Unfortunately, radiation-induced defects in CCDs complicated the mission and made the image processing a difficult task. These defects degraded the quality of the images introducing impulsive noise and possibly also affected the performance of star trackers, which are crucial for the stability of pointing of the spacecraft. 

The chopping mode of observing was introduced at the beginning of 2015 to mitigate the effects of the increasing CCD radiation damage and the influence of CTI on the photometry (see Paper II, Sect.\,6.2.3). The light curves for stars affected by CTI provided by the normal pipeline were contaminated by distinctive artefacts and therefore were scientifically almost useless. A switch to the chopping mode of observing improved the situation significantly, which we have discussed in Sect.\,\ref{normvschop} and have demonstrated using six setups of the UBr data as examples (Fig.\,\ref{chop-norm}). After changing the CCD read-out rate in July 2015, the CTI artefacts practically disappeared. The comparison of results from both pipelines shows the advantage of the chopping pipeline for almost all observations, particularly for those affected by CTI. In addition, it was shown that the advantage of using the chopping pipeline has become greater with time. This is strictly related to the increase of a number of hot pixels generated by radiation in orbit with time. The chopping mode offers much better perspectives for improvements. One possibility is a reduction of the impulsive noise in difference images. There are many techniques that address the problem \citep[see, e.g.,][]{2016MNRAS.463.2172P} and we will search for one that will be optimal for BRITE data. Other planned changes to the photometric pipeline are discussed below. In the context of the unavoidable increase of radiation damage of CCDs with time in all BRITE satellites, it is clear that the chopping mode will allow to extend considerably the length of the mission. 
The development plan of the reduction pipelines is as follows:
\begin{itemize}
\item[$\bullet$]A new pipeline for images obtained in the stare mode of observing will be developed. The images will be re-reduced using thresholded apertures obtained in a similar way as is presently done in the chopping pipeline (see Sect.\,\ref{phopt}). In addition, the same parameters for decorrelation which are now provided with DR5, but are lacking in previous data releases, will be provided. All this should improve the quality of the photometry obtained from these data obtained from the first two years of observations and allow for a better correction for instrumental effects.
\item[$\bullet$]The positions of stars in subrasters in a given frame are not independent because a satellite is a rigid body. We plan to take advantage of this fact to improve precision of the estimation of the positions of stellar centroids.
\item[$\bullet$]Profile fitting photometry will be investigated. This will include temperature dependent variability of intrinsic stellar profiles and the usage of a model of satellite movement during exposures. This movement is the primary cause of smearing of stellar images. Again, the rigidity of a satellite will be used to account for the effect of smearing.
\item[$\bullet$]Several impulsive noise reduction techniques for initial hot pixel replacement will be tested. Currently, a simple median filter algorithm is employed and the hot pixels are assumed to be stable between two consecutive exposures, which is not always true.
\end{itemize}

At present, the BRITE satellites provide photometric data in two passbands at the 1~mmag per orbit level. At the time of writing, the data for over 400 stars have been acquired, often during multiple runs. For about 90\% of these stars images were reduced and extracted photometric data sent to the users.
At the same time, decorrelation techniques have also been developed to account for all identified instrumental effects. The detection threshold for periodic variability in the best datasets reaches 0.15\,--\,0.2~mmag. This means that BRITEs provide the best photometry ever made for many of the brightest stars in the sky. The additional advantage of BRITE photometry is good frequency resolution and small aliasing in the most interesting frequency region (below 10~d$^{-1}$) for stars having observations obtained by two or more satellites. Good frequency resolution is a consequence of the long observing runs lasting 5\,--\,6 months and the fact that some fields are re-observed (see Fig.\,\ref{sat-flds} and Table \ref{rel-flds}). All this shows that precise photometry from low-Earth orbits with the use of nano-satellites, despite the difficulties and the occurrence of instrumental effects, is a scientifically viable undertaking.
\begin{acknowledgements}
This reasearch was supported by the following Polish National Science Centre grants: 2016/21/D/ ST9/00656 (A.\,Po\-po\-wicz), 2016/21/B/ST9/01126 (A.\,Pigulski), 2015/17/N/ST7/03720 (K.\,Ber\-na\-cki), and 2015/18/A/ST9/00578 (E.\,Zoc{\l}o\'nska, G.\,Handler). A.\,F.\,J.\,Moffat, S.\,M.\,Rucinski, and G.\,A.\,Wade are grateful for financial aid from NSERC (Ca\-na\-da). In addition, A.\,F.\,J.\,Moffat acknowledges support from FQRNT (Quebec). O.\,Koudelka, R.\,Kuschnig, and W.\,Weiss acknowledge support by the Austrian Research Agency (FFG) as part of the Austrian Space Application Programme (ASAP). K.\,Zwintz acknowledges support by the Austrian Fonds zur F\"orderung der wissenschaftlichen Forschung (FWF, project V431-NBL). The Polish contribution to the BRITE project is funded by the NCN PMN grant 2011/01/M/ST9/05914. We thank Radek Smolec for his comments made upon reading the manuscript and Monica Chaumont and Daniel Kekez (UTIAS-SFL) for making test images from the integrated UBr satellite available. We thank anonymous referee for the useful comments.
\end{acknowledgements}
\bibliographystyle{aa} 
\bibliography{mybib} 
\begin{appendix}
\section{Data releases and setups}\label{drs}
Following one year of proprietary period, and a possible six month extension for justified cases, BRITE data are made public through the BRITE Public Data Archive\footnote{https://brite.camk.edu.pl/pub/index.html} (BPDA). Presently, the data are available as four data releases, Data Release 2 (DR2\footnote{DR1 (Data Release 1) was later superseded by DR2 in which additional flag indicating if the aperture extends over the subraster edge, has been added. Therefore, DR1 data are no longer in use.}) to Data Release 5 (DR5); see Table \ref{datar}. The information on 15 released fields up to now is summarized in Table \ref{rel-flds}.
\begin{figure*}[!t]
\centering
\includegraphics[width=0.9\textwidth]{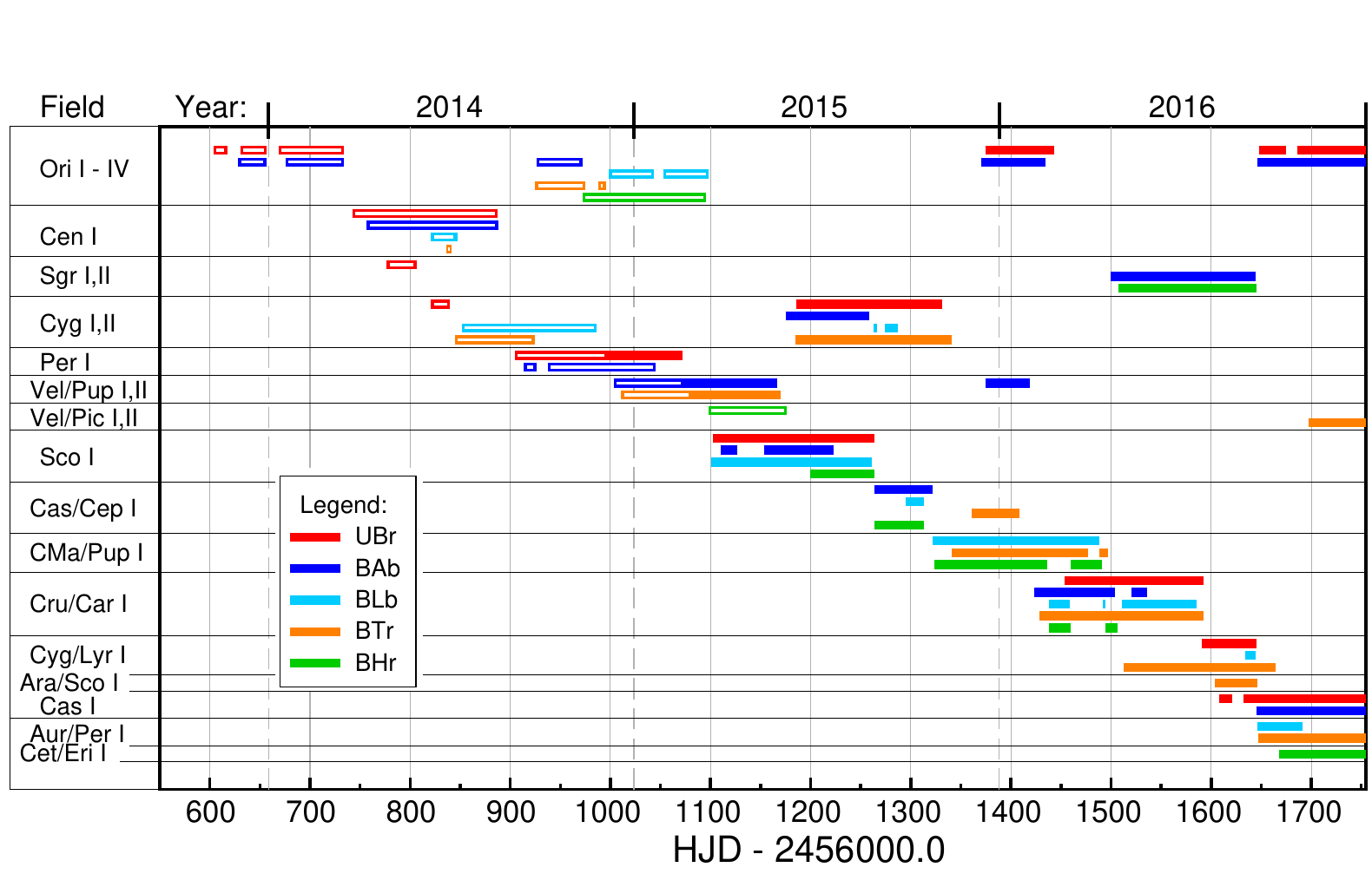}
\caption{Distribution of the observations from all five BRITE satellites untill the end of 2016. The data obtained in the stare and chopping observing modes are shown with unfilled and filled bars, respectively.}
\label{sat-flds}
\end{figure*}
\begin{table*}[!t]
\centering
\small
\caption{Format of text files related to BRITE data releases.}
\begin{tabular}{ccccccccccccc}
\hline\noalign{\smallskip}
Data & \multicolumn{12}{c}{Column in data file}\\
release & (1) & (2) & (3) & (4) & (5) & (6) & (7) & (8) & (9) & (10) & (11) & (12)\\
\noalign{\smallskip}\hline\noalign{\smallskip}
DR2 & HJD & FLUX & XCEN & YCEN & CCDT & JD & FLAG & --- & --- & --- & --- & ---\\
DR3 & HJD & FLUX & XCEN & YCEN & CCDT & JD & PSFC & RTSC(+) & --- & --- & --- & ---\\
DR4 & HJD & FLUX & XCEN & YCEN & CCDT & JD & PSFC1 & PSFC2 & RTSC(+/$-$) & --- & --- & ---\\
DR5 & HJD & FLUX & XCEN & YCEN & CCDT & JD & PSFC1 & PSFC2 & RTSC(+/$-$) & RTSP & APER0 & FLAG \\
\noalign{\smallskip}\hline
\label{datar}
\end{tabular}
\tablefoot{The column designations have the following meaning: HJD -- heliocentric Julian Day of the beginning of exposure; FLUX -- flux resulting from aperture photometry, in ADU; XCEN -- $x$ coordinate of the centroid of the stellar profile in the coordinate frame of a subraster; see Eq.\,(\ref{COG1}), in pixels; YCEN -- the same for the $y$ coordinate; CCDT -- temperature of CCD detector (in $^\circ$C); JD -- Julian Day of the beginning of observation; it can be useful for identifying frames, which are labeled with JD; FLAG -- a 0/1 flag indicating if for a given frame the aperture was fully rendered within the subraster (1) or not (0); PSFC = PSFC1 -- the first blurring parameter, see Appendix \ref{decor}; PSFC2 -- the second blurring parameter; see Appendix \ref{decor}; RTSC -- a parameter related to RTS in columns; see Appendix \ref{decor}, RTSP -- a parameter related to RTS in pixels within the optimal apertures, see Appendix \ref{decor}, APER0 -- median value of background in a difference image.
}
\end{table*}
\begin{table*}[!t]
\centering
\small
\caption{BRITE data distributed to users up to now. Two observing modes were used: stare mode of observing (S) and chopping mode (Ch).}
\begin{tabular}{cccr@{.}lr@{.}lr@{.}lr@{.}lr@{.}l}
\hline\noalign{\smallskip}
      &    Data & Observing  & \multicolumn{10}{c}{Total span [d] / Stars observed / Setup(s)}\\
Field & release & mode & \multicolumn{2}{c}{UBr} & \multicolumn{2}{c}{BAb} & \multicolumn{2}{c}{BLb} & \multicolumn{2}{c}{BTr} & \multicolumn{2}{c}{BHr} \\
\noalign{\smallskip}\hline\noalign{\smallskip}
Ori I & DR2 & S & 130&2 / 15 / 7 & 105&7 / 15 / 3, 4 & \multicolumn{2}{c}{---} & \multicolumn{2}{c}{---} & \multicolumn{2}{c}{---} \\
Cen I & DR2 & S & 145&4 / 30 / 4 & 131&4 / 30 / 4 & 26&7 / 16 / 1 & 6&0 / 31 / 1 & \multicolumn{2}{c}{---}\\
Sgr I & DR2 & S & 29&8 / 19 / 1 & \multicolumn{2}{c}{---} &\multicolumn{2}{c}{---} &\multicolumn{2}{c}{---} &\multicolumn{2}{c}{---}\\
Cyg I & DR2 & S & 19&0 / 24 / 1 & \multicolumn{2}{c}{---} & 135&0 / 22 / 1, 2 & 79&8 / 31 / 1, 2 & \multicolumn{2}{c}{---}\\
Per I & DR2 & S / Ch & 167&9 / 33 / 2, 3, 6, 7 & 131&7 / 19 / 1\,--\,5 & \multicolumn{2}{c}{---} &\multicolumn{2}{c}{---} &\multicolumn{2}{c}{---}\\
Ori II & DR2 & S & \multicolumn{2}{c}{---} & 45&7 / 23 / 2\,--\,4 & 99&8 / 25 / 3, 6&  70&9 / 34 / 1\,--\,3 & 123&3 / 26 / 2, 5\,--\,7\\
Vel / Pup I & DR2/3 & S / Ch & \multicolumn{2}{c}{---} & 162&9 / 31 / 1\,--\,7 & \multicolumn{2}{c}{---} & 159&3 / 36 / 1\,--\,5 & \multicolumn{2}{c}{---}\\
Vel / Pic I & DR2 & S & \multicolumn{2}{c}{---} &\multicolumn{2}{c}{---} &\multicolumn{2}{c}{---} &\multicolumn{2}{c}{---} & 78&3 / 19 / 3\,--\,5\\
Sco I & DR3 & Ch & 161&9 / 18 / 1\,--\,4 & 112&9 / 8 / 2 & 160&6 / 19 / 1\,--\,6 &\multicolumn{2}{c}{---} & 63&8 / 18 / 3 \\
Cyg II & DR3 & Ch & 145&6 / 18 / 2\,--\,4 & 83&3 / 20 / 1\,--\,5 & 24&1 / 25 / 1, 2 & 155&9 / 24 / 1\,--\,5 & \multicolumn{2}{c}{---}\\
Cas / Cep I & DR4 & Ch & \multicolumn{2}{c}{---} & 58&2 / 12 / 1 & 18&2 / 15 / 2\,--\,4 & 47&7 / 7 / 1, 2 & 50&0 / 18 / 1\,--\,3\\
CMa / Pup I & DR4 & Ch & \multicolumn{2}{c}{---} &\multicolumn{2}{c}{---} & 166&5 / 26 / 3\,--\,5 & 156&5 / 21 / 1\,--\,3 & 167&3 / 17 / 3, 4 \\
Ori III & DR4 & Ch & 68&9 / 16 / 1, 2 & 63&3 / 17 / 1, 2 &\multicolumn{2}{c}{---}&\multicolumn{2}{c}{---}&\multicolumn{2}{c}{---}\\
Vel / Pup II & DR4 & Ch & 44&5 / 12 / 1, 2 &\multicolumn{2}{c}{---}&\multicolumn{2}{c}{---}&\multicolumn{2}{c}{---}&\multicolumn{2}{c}{---}\\
Cru / Car I & DR5 & Ch & 138&6 / 21 / 1\,--\,5 & 112&6 / 19 / 1\,--\,6 & 147&3 / 19 / 1\,--\,6 & 163&9 / 27 / 1\,--\,4 & 68&4 / 18 / 1\\
Sgr II & DR5 & Ch & \multicolumn{2}{c}{---} & 144&5 / 12 / 1, 2 & \multicolumn{2}{c}{---} & \multicolumn{2}{c}{---} & 145&7 / 19 / 1\,--\,4\\
Cyg / Lyr I & DR5 & Ch & 54&5 / 7 / 1, 2 &  \multicolumn{2}{c}{---} & 11&5 / 7 / 2, 3 & 152&6 / 18 / 2\,--\,5 & \multicolumn{2}{c}{---}\\
\noalign{\smallskip}\hline
\label{rel-flds}
\end{tabular}
\end{table*}

BRITE-Constellation data are made available as text files with several columns preceded by a header in which some general information on the satellite, data length, data format, etc, is included. The data format for each data release is shown in Table \ref{datar}; the meaning of the parameters included in text files is explained in the table's footnote and Appendix \ref{decor}.

\subsection{Setups}
Data obtained with a given BRITE satellite during its run in a given field are split into parts, so-called setups, during the reduction process. The reason for this splitting is usually a change in observing conditions. When observations of a new field start, stars are positioned only in preliminary subrasters. After a short time, typically one or two days since the onset of observations, the subrasters need to be readjusted because some stars are not properly placed (e.g.~too close to a subraster edge). The other reason for an adjustment could be a change of exposure time. Consequently, the first setup of a given run is usually short. The subsequent changes to the observing procedure also result in new setups. The changes can be the following:
\begin{itemize}
\item[$\bullet$] Changes to the list of observed stars. Sometimes, especially if there were data transmission issues, the list of observed stars was shortened to allow complete transmission of the data. Next, some stars were dropped because of their faintness. On the other hand, it might happen that the list of observed stars was extended to include more objects, usually faint, because both data transmission and the overall performance of a satellite warranted good photometry for them. This kind of change usually did not affect stars that were observed throughout the run and therefore for such stars the two setups, before and after the change, can be merged without applying offsets. 
\item[$\bullet$] Long term offsets in `absolute' pointing drove stars close to the edge of the subrasters. The reason for this small loss of collimation between tracker and the main camera is not known. This required adjustments of positions of all subrasters. 
\item[$\bullet$] Due to the limitations of the hardware used for reduction, images from the longest setups were split in two parts. This means that for these two parts there was no change in the observing conditions, but because the data were reduced independently, the two parts may show slightly different dependencies on parameters as a consequence of a possible difference in the shape and size of the optimal apertures. In such situations, the two parts have the same number of the setup, but are marked as `part1', and `part2'.
\end{itemize}

\section{Parameters for decorrelation}\label{decor}
The raw BRITE photometry is subject to many instrumental effects. These effects occur as a consequence of: (i) lack of flat-fielding, (ii) lack of cooling of the detector, (iii) temperature effects on telescope optics, (iv) non-perfect stability of a satellite during exposure, (v) other effects. Fortunately, most of these effects can be to a large extent mitigated by decorrelating raw magnitudes with a number of parameters which are provided together with the raw fluxes. Examples of decorrelation applicable for BRITE data were presented e.g.~by \cite{2016A&A...588A..55P} and \cite{2017arXiv170400576B}. In general, the presented methods iteratively determine the dependencies between decorrelation parameters and raw magnitudes. If large-amplitude intrinsic variability is present, the variability is modelled by a superposition of sinusoidal terms and subtracted. Then, decorrelations are performed using residuals from the fit, but applied to the original light curve. The fitting of the model is a part of the iterative decorrelation sequence. For more details the interested reader is kindly referred to the original papers.

Decorrelations are now regarded as the most important preliminary step in the analysis of BRITE data. In most cases only after this step do the data become scientifically useful. Depending on the data release, different sets of parameters are provided (Table \ref{datar}). They are described in the following.

The XCEN, YCEN and CCDT parameters are provided in all data releases. The XCEN and YCEN are, respectively, the $x$ and $y$ coordinates of the centroid of the stellar profile in the coordinate system of a subraster. The way in which they were obtained is described in detail in Sects~\ref{photo} and \ref{autclass}. CCDT is the average temperature of the CCD, measured by four temperature sensors; see Fig.\,3 in Paper II.

The FLAG parameter, provided first in DR2, indicates if the optimal aperture is fully rendered within a subraster (FLAG $=$ 1) or not (FLAG $=$ 0). This parameter was abandoned in DR3 and DR4 because images which would have FLAG $=$ 0 were eliminated at the classification step in the reduction of chopping data. This was changed back in the chopping pipeline for DR5. The images with the aperture extending beyond the subraster edge are not used in the calculation of the optimal aperture, but the photometry is done and it is the decision of the user whether to reject flagged data points or not. For this purpose, the data points obtained from such images are flagged with FLAG $=$ 0.

It was recognized very early that due to the imperfect stability of the satellites, BRITE images are slightly blurred. The amount of blurring has a direct consequence on the flux measured through a constant aperture. Therefore, starting from DR3, i.e.~the first data release for chopping data, the first parameter which is a good measure of blurring, PSFC, was introduced. Later on, since DR4, another blurring parameter, PSFC2, was added. These two parameters are strongly correlated but using both in decorrelations allows for a better correction for the effects of blurring. The first parameter, PSFC (in DR3) or PSFC1 (in DR4 and DR5), is defined as follows: 
\begin{equation}
{\rm PSFC} = {\rm PSFC1} = \sum_{{\rm p} \in M_{+}}{\left(\frac{I_{\rm p}}{I_{\rm total}}\right)^2},\quad I_{\rm total} = \sum_{{\rm p} \in M_{+}}{I_{\rm p}},
\end{equation}
where $I_{\rm p}$ is the signal in pixel $p$ belonging to mask $M_+$ (see Sect.\,\ref{autclass} for the explanation of the concept of masks),  and $I_{\rm total}$ is the sum of signals from all pixels within the mask $M_+$. Since the signal in each pixel is normalized by $I_{\rm total}$, PSFC is independent of variations of stellar flux (variability of a star). When the stability of a satellite is good, the relative intensities of pixels reach their maximum values, and therefore their squares are maximized. When blurring occurs, the charge is spread more evenly among pixels and the value of PSFC is reduced. The idea of this parameter is based on the image energy (the sum of the squared intensities over all pixels), which is largest when the image is sharp and lower when the flux is spread over a larger CCD area.

The other blurring parameter, PSFC2, introduced with DR4, is defined as
\begin{equation}
{\rm PSFC2} ={\mbox{1}\over{\mbox{4}I_R^2}}\sum_{k=1}^4 \sum_{i=1}^{L} I_i J_{i,k},\quad I_R = \sum_{{\rm p} \in R} I_p,
\end{equation}
where $I_R$ is the sum of signals from all pixels in a subraster $R$, $J_{i,k}$ is the signal in the $i$-th pixel in the image that has been shifted by one pixel in one of the four ($k$) directions: up, down, left and right. The summation is performed over $L$ pixels, which the image $I$ has in common with its shifted copy $J$. This smearing parameter is based on the correlation of the image with its shifted copies. The correlation increases when smearing is larger. Thus, sharp images will have low values of PSFC2. The PSFC1 and PSFC2 parameters are the same as parameters A and B described by \cite{2016A&A...588A..55P} in Appendix A of their paper.

Starting from DR3, a parameter related to the RTS phenomenon was added to the data files. The RTSC parameter is the indicator of a possible RTS in a subraster column. The dark current RTS occurs rarely in a single pixel within the optimal aperture and if this is the case, we get a significant outlier in the final light curve, which can be easily removed. An RTS that affects a whole column results in a relatively smaller photometric offset, due to the much lower dark charge in columns. If the RTS in a column occurs, the difference image contains dimmer or brighter vertical line (depending on the direction of the transition). For example, in the upper right image in Fig.\,\ref{bad_images_CTI} a slightly a dimmer line runs through the negative stellar profile. As such a disturbance may not be properly identified and rejected, the RTSC parameter was introduced to deal with it. If a frame is not affected by a column RTS, the RTSC parameter is equal to several ADUs. The RTS phenomena change it to several tens of ADUs. In order to calculate the RTSC parameter, the median intensity is computed in each column of a difference image, excluding pixels identified as belonging to stellar profile. In DR3 the RTSC was defined as the maximum of the absolute values of these medians. Since DR4 the sign is preserved, so that RTSC can be either positive or negative. In order to indicate this small change in the definition of RTSC, the RTSC in DR3 is marked as RTSC(+) in Table \ref{datar}, denoting that only non-negative values are possible, while for DR4 and DR5, it is marked as RTSC(+/$-$), showing that both negative and positive values are allowed.

Finally, with DR5 we have introduced two new parameters, RTSP, and APER0 (see Table \ref{datar}). RTSP is related to the RTS phenomenon too, but this time, the signals in all pixels within the optimal aperture are compared. The analysis for a given $i$-th image is made using two neighbouring raw images, $(i-1)$-th and $(i+1)$-th. Since in these two images the star is located at the opposite side of a subraster compared to the $i$-th image, it is easy to find the differences in intensity between the pixels in the $i-1$-th and $i+1$-th images that are confined in the stellar profile in the $i$-th image. The RTSP is defined as the maximum of the absolute values of these differences. Similarly to RTSC, its sign is preserved, so that RTSC can be either positive or negative. The APER0 parameter is the median value of all pixels located outside both apertures in a difference image. It was introduced to account for small differences in background which may occur between two positions in the chopping mode of observing. The differences can be introduced by the bright Moon and/or Earth if not far from the observed field.
\end{appendix}
\end{document}